\documentclass[12pt]{article}
\usepackage{amssymb}
\input epsf

\makeatletter
\def\numberbysection{\@addtoreset{equation}{section}
 	\def\theequation{\thesection.\arabic{equation}}}
\makeatother

\numberbysection

\newcommand{\be}{\begin{eqnarray}}
\newcommand{\ee}{\end{eqnarray}}
\newcommand{\non}{\nonumber}
\newcommand{\tr}{\mathop{\rm tr}\nolimits}
\newcommand{\id}{\mathbb{I}}
\newcommand{\M}{\ensuremath{\mathsf{M}}}
\newcommand{\R}{\ensuremath{\mathsf{R}}}
\newcommand{\SSS}{\ensuremath{\mathsf{S}}}
\newcommand{\X}{\ensuremath{\mathsf{X}}}
\newcommand{\Y}{\ensuremath{\mathsf{Y}}}
\newcommand{\Z}{\ensuremath{\mathsf{Z}}}
\newcommand{\aaa}{\ensuremath{\mathfrak{a}}}
\newcommand{\bbb}{\ensuremath{\mathfrak{b}}}
\newcommand{\ccc}{\ensuremath{\mathfrak{c}}}
\newcommand{\ddd}{\ensuremath{\mathfrak{d}}}
\newcommand{\ttt}{\ensuremath{\mathfrak{t}}}
\newcommand{\LL}{\ensuremath{\mathfrak{L}}}
\newcommand{\LLL}{\ensuremath{\Lambda}}

\begin{document}

\begin{titlepage}
\strut\hfill UMTG--224
\vspace{.5in}
\begin{center}

\LARGE Exact solution of the supersymmetric sinh-Gordon model with 
boundary\\[1.0in]
\large Changrim Ahn\footnote{Department of Physics, Ewha Womans 
University, Seoul 120-750, South Korea} and Rafael I. Nepomechie\footnote{Physics Department, 
P.O. Box 248046, University of Miami, Coral Gables, FL 33124 USA}\\

\end{center}

\vspace{.5in}

\begin{abstract}
The boundary supersymmetric sinh-Gordon model is an integrable 
quantum field theory in $1+1$ dimensions with bulk $N=1$ 
supersymmetry, whose bulk and boundary $S$ matrices are not diagonal. 
We present an exact solution of this model. In particular, we derive 
an exact inversion identity and the corresponding thermodynamic Bethe 
Ansatz equations. We also compute the boundary entropy, and find a 
rich pattern of boundary roaming trajectories corresponding to $c < 3/2$ 
superconformal models.
\end{abstract}

\end{titlepage}

\setcounter{footnote}{0}

\section{Introduction}

The supersymmetric sinh-Gordon (SShG) model \cite{DF}-\cite{Ah1} is one of 
the simplest examples of a  $1+1$ dimensional integrable 
quantum field theory with $N=1$ supersymmetry. Indeed, the particle 
spectrum consists of one Boson and one Fermion which have equal mass 
and which enjoy factorized scattering \cite{ZZ1}. As such, SShG is a 
valuable toy model.

In this article, we consider the boundary SShG model, with boundary 
conditions that preserve the bulk integrability, but not necessarily 
the bulk supersymmetry \cite{GZ}-\cite{MS1}. In addition to its usefulness 
as a simple prototype, we expect that this model may also have 
applications to quantum impurity problems \cite{Sa}.

An interesting feature of the boundary SShG model is that the $S$
matrices which have been conjectured for both bulk and boundary
scattering are not diagonal.  Our main objective is to perform a
thermodynamic Bethe Ansatz (TBA) analysis \cite{Ah1}, \cite{YY}-\cite{LMSS}
for this model, using these conjectured $S$ matrices as
inputs.  Such analysis can provide checks on the input $S$ matrix
data, as well as information about the underlying boundary conformal
field theory \cite{BPZ}-\cite{Ca}.

Conventional wisdom suggests that the problem of determining the 
necessary Bethe Ansatz equations is intractable, due to the fact that 
the boundary $S$ matrix is not diagonal. Nevertheless, we succeed to 
determine the Bethe Ansatz equations and carry out the TBA analysis 
for the boundary SShG model. This is the first example of a model 
with non-diagonal boundary $S$ matrix which is exactly solved.

We also obtain an expression for the boundary entropy
\cite{AL},\cite{LMSS} for the boundary SShG model.  Moreover, we find
a rich pattern of boundary roaming trajectories corresponding to 
$c < 3/2$ superconformal models \cite{BKT}, \cite{FQS}, thereby
generalizing previous work on bulk \cite{Za4}-\cite{DR} and boundary
\cite{LSS}, \cite{AR} roaming.

The outline of this article is as follows.  In Sec.  2 we review the
scattering theory of the boundary SShG model, which serves as our
input.  Here we also show that the strong-weak duality symmetry of the
bulk model (see, e.g., \cite{Mu}) extends also to the model with boundary. 
Moreover, we introduce the notations and conventions which are used
throughout the paper.  In Sec.  3 we formulate the so-called Yang
matrix \cite{Ya} and relate it to a commuting transfer matrix, which
is the true starting point of any TBA analysis.  For the problem at
hand, we require a boundary version of the Yang matrix \cite{FS},
\cite{GMN}, which presents an interesting complication with respect to
the more familiar case of periodic boundary conditions.  In Sec.  4 we
use the open-chain fusion formula \cite{MN} to derive an exact
inversion identity, using which we obtain the eigenvalues of the
transfer matrix in terms of roots of certain Bethe Ansatz equations. 
That such an inversion identity exists is presumably due to the fact
that the bulk $S$ matrix satisfies the so-called free Fermion
condition \cite{FW}, \cite{Fe}.  In Sec.  5 we use these results to
derive the TBA equations.  Certain remarkable identities lead to very
simple formulas, in particular for the boundary entropy.  In Sec.  6
we use our result for the boundary entropy to obtain boundary roaming
trajectories.  Finally, in Sec.  7 we discuss our results and describe
some possible generalizations.

\section{Review of boundary SShG scattering theory}

In this Section, we review the bulk and boundary $S$ matrices which
have been proposed for the boundary supersymmetric sinh-Gordon model. 
As mentioned in the Introduction, these $S$ matrices will be used as
inputs in the calculations that follow.  We also show that the
strong-weak duality symmetry of the bulk model extends also to the
model with boundary.

\subsection{Bulk}

In order to understand the SShG scattering theory, it is essential to 
first consider a related model, namely, the supersymmetric 
sine-Gordon (SSG) model, whose Euclidean-space Lagrangian density is 
given by
\be
{\cal L}^{SSG}_{bulk}= {1\over 2}
\left( \partial_{z}\phi \partial_{\bar z}\phi 
+ \bar \psi \partial_{z}\bar \psi
+ \psi \partial_{\bar z}\psi \right) 
+ i M \bar \psi \psi \cos \beta \phi
+ {M^{2}\over 2 \beta^{2}}\sin^{2} \beta \phi \,,
\label{SSGbulkLagrangian}
\ee
where $\phi(z \,, \bar z)$ is a real scalar field, $\psi(z \,, \bar
z)$ and $\bar \psi(z \,, \bar z)$ are the components of a Majorana
spinor field, and $\beta$ is the dimensionless coupling constant.  The
Lagrangian density for the supersymmetric sinh-Gordon (SShG) model is
obtained by analytic continuation to imaginary coupling, i.e. setting
$\beta = i \hat \beta$ with $\hat \beta$ real:
\be
{\cal L}^{SShG}_{bulk}= {1\over 2}
\left( \partial_{z}\phi \partial_{\bar z}\phi 
+ \bar \psi \partial_{z}\bar \psi
+ \psi \partial_{\bar z}\psi \right) 
+ i M \bar \psi \psi \cosh \hat \beta \phi
+ {M^{2}\over 2 \hat \beta^{2}}\sinh^{2} \hat \beta \phi \,.
\label{SShGbulkLagrangian}
\ee
Both of these models have $N=1$ supersymmetry (without topological 
charge) \cite{DF}, \cite{Hr} and are integrable \cite{FGS}, 
\cite{GS}. \footnote{Of course, a similar relation exists between the 
usual (non-supersymmetric) sine-Gordon (SG) and sinh-Gordon (ShG) 
models. It is more straightforward to infer the scattering theory for 
the trigonometric (SG, SSG) models than for the hyperbolic (ShG, 
SShG) models, because the former have kinks with topological charge, 
whose non-diagonal $S$ matrices must satisfy highly restrictive 
constraints \cite{ZZ1},\cite{ABL},\cite{BL}. The $S$ matrices for 
the hyperbolic models are inferred by analytic continuation of the 
corresponding breather $S$ matrices, as is explained in more detail 
below.}

Observe that SSG has a periodic potential, which admits classical 
soliton solutions that interpolate between neighboring minima. 
Correspondingly, it has been proposed \cite{ABL}-\cite{Ah2} that the 
SSG quantum spectrum consists of supersymmetric multiplets of kinks of 
mass $m$ and breathers (bound states of kinks) of mass $m_{n} = 2 m 
\sin( n \alpha \pi)$,  $n = 1 \,, 2\,, \ldots\,,  [{1\over 2 \alpha}]$, 
where
\be
\alpha = {{\beta^{2}\over 4 \pi}\over 1 - {\beta^{2}\over 4 \pi}} \,,
\label{alpha}
\ee
and $[x]$ denotes integer part of $x$.  Hence, breathers can be
present only if $0 < \alpha < {1\over 2}$.  The lightest ($n=1$)
breathers are identified as the elementary particles (Boson, Fermion)
corresponding to the fields in the Lagrangian density
(\ref{SSGbulkLagrangian}).

Upon making the analytic continuation to SShG (which is the model of
primary interest), we see that the potential is no longer periodic,
and hence, there are no longer any classical soliton solutions.  Thus,
the SShG quantum spectrum does not contain kinks; it consists only of
the elementary particles of some mass $m$ corresponding to the fields
in the Lagrangian density (\ref{SShGbulkLagrangian}), i.e.
corresponding to the $n=1$ SSG breather.  Setting $\beta = i \hat
\beta$ in Eq.  (\ref{alpha}), we obtain
\be
\alpha = 
- {{\hat \beta^{2}\over 4 \pi}\over 1 + {\hat \beta^{2}\over 4 \pi}} 
\equiv -B \,,
\label{B}
\ee
where we have introduced the SShG parameter $B$.

Since the SShG spectrum corresponds to the $n=1$ SSG breather, we infer 
that the SShG $S$ matrix $S(\theta)$ for two particles of rapidities 
$\theta_{1} \,, \theta_{2}$ (and corresponding energy 
$E_{i} = m \cosh \theta_{i}$ and momentum $P_{i} = m \sinh \theta_{i}$, 
$i = 1 \,, 2$) is given by the analytic continuation of the $n=1$ SSG 
breather $S$ matrix,
\be
S(\theta) = S_{ShG}(\theta)\ S_{SUSY}(\theta) \,,
\label{bulkSmatrix}
\ee
where $\theta=\theta_{1}-\theta_{2}$.  The scalar factor
$S_{ShG}(\theta)$ is given by
\be
S_{ShG}(\theta) = {\sinh \theta - i \sin( 2 B \pi)\over
 \sinh \theta + i \sin( 2 B \pi)} \,.
\label{bulkShG}
\ee
This is the $S$ matrix of the usual (non-supersymmetric) sinh-Gordon
model \cite{VG},\cite{STW}, which is the analytic continuation of the
$n=1$ SG breather $S$ matrix \cite{ArKo}, but with a different
dependence on the coupling constant.  It satisfies
\be
S_{ShG}(\theta)\ S_{ShG}(-\theta) = 1 \,, 
\qquad S_{ShG}(\theta) = S_{ShG}(i\pi - \theta) \,. 
\label{ShGproperties}
\ee 

The factor $S_{SUSY}(\theta)$ is given
by \footnote{The matrix $S_{SUSY}(\theta)$ for SSG was first obtained
\cite{SW} in terms of an unknown parameter $\Delta$ by solving the
constraints coming from supersymmetry and factorization.  The
identification of $\Delta$ in terms of $\alpha$ was made in
\cite{Ah2}.}
\be
S_{SUSY}(\theta) = Y(\theta)\ R(\theta) \,,
\label{bulkSUSY}
\ee
where $R(\theta)$ is a $4 \times 4$ matrix acting on the tensor 
product space $V \otimes V$, where $V$ is the 2-dimensional vector 
space of 1-particle states. We choose $\{ |b(\theta) \rangle \,, 
|f(\theta) \rangle \}$ to be the basis of $V$ (corresponding to a 
Boson, Fermion with rapidity $\theta$, respectively); and hence, the 
basis of $V \otimes V$ is given by $\{ |b_{1} \,, b_{2} \rangle \,, 
|b_{1} \,, f_{2} \rangle \,, |f_{1} \,, b_{2} \rangle \,, 
|f_{1} \,, f_{2} \rangle \}$, where  $|b_{1} \,, b_{2} \rangle 
\equiv |b(\theta_{1}) \,, b(\theta_{2})  \rangle$ , etc. In this 
basis, $R(\theta)$ is given by
\be
R(\theta) = \left( \begin{array}{cccc}
	a_{+}(\theta) &0         &0           &d(\theta)    \\
        0             &b         &c(\theta)   &0            \\
	0             &c(\theta) &b           &0            \\
	d(\theta)     &0         &0           &a_{-}(\theta)
\end{array} \right) \,, 
\label{bulkRmatrix}
\ee 
with
\be
a_{\pm}(\theta) = \pm 1 -  {2 i\sin B \pi\over \sinh \theta} \,, \qquad
b = 1 \,, \qquad c = - {i\sin B \pi\over \sinh {\theta\over 2}} \,, \qquad
d = -{\sin B \pi\over \cosh {\theta\over 2}} \,.
\label{Rmatrixelements}
\ee
It is important to note that the matrix elements of $R(\theta)$ 
satisfy the ``free Fermion'' condition \cite{FW}, \cite{Ah1}
\be
a_{+}a_{-} + b^{2} = c^{2} + d^{2} \,.
\label{freeFermion}
\ee
The matrix $R(\theta)$ is a solution of the Yang-Baxter equation 
\footnote{\label{convention}We use the very useful convention 
(which is standard in the spin-chain literature \cite{KS} - \cite{Ne}, 
but unfortunately not in the field-theory literature), 
whereby $R_{ij}(\theta)$ acts nontrivially on the 
$i^{th}$ and $j^{th}$ vector spaces. For instance, in the Yang-Baxter 
equation, the $R$ matrices act on $V^{\otimes 3}$, and therefore
$R_{12}(\theta)=R(\theta)\otimes \id \,, \quad R_{23}(\theta)=\id \otimes 
R(\theta)$, etc., where $\id$ is the unit matrix.}
\be
R_{12}(\theta_{1}-\theta_{2})\ R_{13}(\theta_{1}-\theta_{3})\ 
R_{23}(\theta_{2}-\theta_{3})\ =  
R_{23}(\theta_{2}-\theta_{3})\ R_{13}(\theta_{1}-\theta_{3})\ 
R_{12}(\theta_{1}-\theta_{2}) \,.
\label{YangBaxter}
\ee
This matrix is both $\sf P$ and $\sf T$ invariant,
\be
{\cal P}_{12}\ R_{12}(\theta)\ {\cal P}_{12} = R_{12}(\theta) \,,
\qquad
R_{12}(\theta)^{t_{1} t_{2}} = R_{12}(\theta) \,, 
\label{Rsymmetries}
\ee
where $t_{i}$ denotes transposition in the $i^{th}$ space, and 
${\cal P}$ is the permutation matrix
\be
{\cal P} = \left( \begin{array}{cccc}
 1  &0  &0  &0  \\ 
 0  &0  &1  &0  \\
 0  &1  &0  &0  \\
 0  &0  &0  &1 
\end{array} \right) \,.
\label{permutation}
\ee 

Moreover, $Y(\theta)$ is a scalar factor given by
\be
Y(\theta) = {\sinh{\theta\over 2}\over \sinh{\theta\over 2} 
- i \sin B \pi} \exp \left( - \int_{0}^{\infty} {dt\over t}\
{\sinh (i t \theta/\pi) \sinh(t(1+B)) \sinh (t B)\over \cosh t 
\cosh^{2}{t\over 2}} \right) \,,
\label{bulkYfactor}
\ee
which is a solution of the unitarity and crossing constraints
\be
Y(\theta)\ Y(-\theta) = {\sinh^{2}{\theta\over 2}\over \sinh^{2}{\theta\over 2} 
+ \sin^{2} B \pi} \,, \qquad Y(\theta) = Y(i\pi -\theta) \,.
\label{bulkYproperties}
\ee

Let us denote the total scalar factor by $Z(\theta)$
\be
Z(\theta) = S_{ShG}(\theta)\ Y(\theta) \,.
\label{bulkZ}
\ee
One can show that $Z(\theta)$ has the integral representation
\cite{Ah1}
\be
Z(\theta) = {\sinh{\theta\over 2}\over \sinh{\theta\over 2} 
+ i \sin B \pi} \exp \left( \int_{0}^{\infty} {dt\over t}\
{\sinh (i t \theta/\pi) \sinh(t(1-B)) \sinh (t B)\over \cosh t 
\cosh^{2}{t\over 2}} \right) \,,
\label{bulkZfactor}
\ee
which is the same as the expression in Eq. (\ref{bulkYfactor}), except 
with $B \rightarrow -B$.  It has no poles \footnote{Although
$Y(\theta)$ has a pole at $\theta = i 2 B \pi$, it is canceled by a
corresponding zero of $ S_{ShG}(\theta)$.} in 
the physical strip ($0 < Im \ \theta < \pi$), provided $B$ lies in the 
range
\be
0 < B < 1 \,,
\label{Brange}
\ee 
which corresponds to $0 < \hat \beta^{2} < \infty$.

In short, the proposed SShG bulk $S$ matrix $S(\theta)$ is given by
\be
S(\theta) = Z(\theta)\ R(\theta) \,,
\label{bulkS}
\ee
where the scalar factor $Z(\theta)$ is given by Eq. 
(\ref{bulkZfactor}) and the matrix $R(\theta)$ is given by Eqs. 
(\ref{bulkRmatrix}), (\ref{Rmatrixelements}).

It is known (see, e.g., \cite{Mu}) that the SShG bulk $S$ matrix is
invariant under the strong-weak duality transformation $\hat \beta
\rightarrow 4\pi/\hat \beta$, which implies
\be
B \rightarrow 1 - B \,.
\label{duality}
\ee
Indeed, this invariance can be checked by inspection of the matrix 
elements (\ref{Rmatrixelements}) of $R(\theta)$ and the expression
(\ref{bulkZfactor}) for $Z(\theta)$. (The factors $S_{ShG}(\theta)$ 
and $Y(\theta)$ are not separately invariant.) Note that this 
transformation maps the range (\ref{Brange}) into itself.

\subsection{Boundary}

We turn now to boundary conditions and boundary scattering, following 
the framework developed by Ghoshal and Zamolodchikov \cite{GZ}. An 
investigation of which boundary terms can be added to the bulk SShG 
model (\ref{SShGbulkLagrangian}) without spoiling (classical) 
integrability has led to the following results \cite{IOZ}: the 
boundary Lagrangian
\be
L^{SShG}_{boundary} = \Lambda \cosh \hat \beta (\phi - \phi_{0})
+ \M \bar \psi \psi + \epsilon \psi + \bar \epsilon \bar \psi  \,,
\qquad \M \ne \pm 1 \,, 
\label{boundaryLagrangian}
\ee
breaks supersymmetry but preserves integrability; and
\be
L^{SShG}_{boundary} = \mp {M\over \hat \beta^{2}} \cosh \hat \beta 
\phi \pm \bar \psi \psi
\ee
preserves both supersymmetry and integrability. Notice that the 
boundary terms (\ref{boundaryLagrangian}) involve a total of 5 
boundary parameters $\Lambda \,, \phi_{0} \,, \M \,, \epsilon \,, 
\bar\epsilon$. If  $\epsilon \,, \bar\epsilon$ are nonzero, then 
Fermion number is not conserved.

The proposed boundary $S$ matrix $\SSS(\theta)$ for a particle of 
rapidity $\theta$ is given by (compare with 
Eq. (\ref{bulkSmatrix})) \footnote{We 
make an effort to distinguish boundary quantities from the 
corresponding bulk quantities by using sans serif letters to denote 
the former, and Roman letters to denote the latter.},\footnote{The 
boundary $S$ matrix is obtained in terms of a set of boundary 
parameters $( \eta \,, \vartheta \,, \varphi \,, \varepsilon )$ by 
solving the boundary Yang-Baxter equation. The relation of these 
parameters to those in $L^{SShG}_{boundary}$ 
(\ref{boundaryLagrangian}) is not known.}
\be
\SSS(\theta)= \SSS_{ShG}(\theta \,; \eta \,, \vartheta )\ 
\SSS_{SUSY}^{(\varepsilon)}(\theta \,; \varphi ) \,.
\label{boundarySmatrix}
\ee
The scalar factor $\SSS_{ShG}(\theta \,; \eta \,, \vartheta )$, which
depends on two boundary parameters $\eta \,, \vartheta$, is given by
\be
\SSS_{ShG}(\theta \,; \eta \,, \vartheta ) = \X_{0}(\theta)\ 
\X_{1}(\theta \,; {4 \eta B\over \pi})\
\X_{1}(\theta \,; {4 i \vartheta B\over \pi}) \,,
\label{boundaryShG}
\ee
where
\be 
\X_{0}(\theta) = (1)(1+2B)(2-2B) \,, \qquad 
\X_{1}(\theta \,; F) = {1\over (1-F)(1+F)} \,,
\ee
with
\be 
(x) \equiv {\sinh({\theta\over 2} +  {i\pi x \over 4})\over
\sinh({\theta\over 2} -  {i\pi x \over 4})} \,.
\ee
This is the boundary $S$ matrix of the usual (non-supersymmetric)
boundary sinh-Gordon model, which is the analytic continuation of the
$n=1$ boundary sine-Gordon breather $S$ matrix \cite{Gh}.  It
satisfies
\be
\SSS_{ShG}(\theta \,; \eta \,, \vartheta )\
\SSS_{ShG}(-\theta \,; \eta \,, \vartheta ) = 1 \,, \quad 
\SSS_{ShG}({i \pi\over 2}+\theta \,; \eta \,, \vartheta )\ S_{ShG}(2\theta)\
= \SSS_{ShG}({i \pi\over 2}-\theta \,; \eta \,, \vartheta ) \,. \non \\
\label{boundShGproperties}
\ee 

The factor $\SSS_{SUSY}^{(\varepsilon)}(\theta \,; \varphi )$ is given 
by \cite{AK}
\be
\SSS_{SUSY}^{(\varepsilon)}(\theta \,; \varphi ) 
= \Y^{(\varepsilon)}(\theta \,; \varphi )\ 
\R^{(\varepsilon)}(\theta \,; \varphi ) \,, 
\label{boundarySUSY}
\ee 
where $\varepsilon$ is a discrete parameter which can be either $+1$ 
or $-1$, and $\varphi$ is a continuous boundary parameter.
$\R^{(\varepsilon)}(\theta \,; \varphi )$ is a $2 \times 2$ matrix 
acting on the vector space $V$ of 1-particle states, which is given by
\be
\R^{(\varepsilon)}(\theta \,; \varphi ) = \left( \begin{array}{cc}
\cosh{\theta\over 2}\ G_{+}^{(\varepsilon)}
+ i \sinh{\theta\over 2}\ G_{-}^{(\varepsilon)}
&  -\varepsilon i \sinh \theta   \\
-i \sinh \theta 
& \cosh{\theta\over 2}\ G_{+}^{(\varepsilon)}
- i \sinh{\theta\over 2}\ G_{-}^{(\varepsilon)}
\end{array} \right) \,, 
\label{boundaryRmatrix}
\ee 
where
\be
G_{\varepsilon'}^{(\varepsilon)} = 
\left\{ \begin{array}{r@{\quad \mbox{if} \quad}l}
r \left( \cosh\varphi + e^{\varepsilon 
\varphi}{\sinh^{2}{\theta\over 2}\over 1 + \varepsilon \sin B \pi} 
\right)
& \varepsilon' = \varepsilon \\
r \left( \sinh\varphi + \varepsilon e^{\varepsilon 
\varphi}{\sinh^{2}{\theta\over 2}\over 1 + \varepsilon \sin B \pi}
\right)
& \varepsilon' = -\varepsilon
\end{array} \right. \,, 
\ee
and 
\be
r = \left({2 (\varepsilon + \sin B \pi)\over \sin B \pi} 
\right)^{1\over 2} \,.
\ee
The matrix $\R^{(\varepsilon)}(\theta \,; \varphi )$ is a solution 
of the boundary Yang-Baxter equation \cite{Ch}
\be
\lefteqn{R_{12}(\theta_{1}-\theta_{2})\ 
\R_{1}^{(\varepsilon)}(\theta_{1}\,; \varphi)\ 
R_{12}(\theta_{1}+\theta_{2})\ \R_{2}^{(\varepsilon)}(\theta_{2}\,; \varphi)} 
\non \\
& & = \R_{2}^{(\varepsilon)}(\theta_{2}\,; \varphi)\ 
R_{12}(\theta_{1}+\theta_{2})\ 
\R_{1}^{(\varepsilon)}(\theta_{1}\,; \varphi)\ R_{12}(\theta_{1}-\theta_{2}) \,.
\label{boundaryYangBaxter}
\ee 
We remark that for $\varphi \rightarrow \pm \infty$, the matrix
$\R^{(\varepsilon)}(\theta \,; \varphi )$ becomes diagonal and 
commutes with linear combinations $Q \pm \bar Q$ of the supersymmetry 
charges \cite{AK},\cite{MS1}.

In order to determine the scalar factor 
$\Y^{(\varepsilon)}(\theta \,; \varphi )$, we recall that the full
boundary $S$ matrix must satisfy boundary unitarity 
$\SSS(\theta)\ \SSS(-\theta) = \id$ and boundary cross-unitarity
\cite{GZ}, which can be written in matrix form as
\be
\tr_{0} \SSS_{0}( {i\pi\over 2} + \theta)^{t_{0}}\ {\cal 
P}_{01}\ S_{01}(2 \theta)^{t_{1}} = 
\SSS_{1}( {i\pi\over 2} - \theta) \,.
\label{boundarycrossing}
\ee
We observe that the matrix $\R^{(\varepsilon)}(\theta \,; \varphi )$ 
satisfies
\be
\R^{(\varepsilon)}(\theta \,; \varphi)\ 
\R^{(\varepsilon)}(-\theta \,; \varphi) = h(\theta)\ \id \,, 
\ee
where
\be
h(\theta) = \left( c_{0} + c_{1} \sinh^{2}{\theta\over 2} +
c_{2} \sinh^{4}{\theta\over 2} \right) \cosh \theta \,, 
\label{h}
\ee 
and
\be
c_{0} = \left\{ \begin{array}{cc}
r^{2} \cosh^{2} \varphi  & \mbox{ if } \varepsilon = + 1\\
r^{2} \sinh^{2} \varphi  & \mbox{ if } \varepsilon = - 1
\end{array} \right. \,, \quad 
c_{1} ={r^{2} e^{\varepsilon 2 \varphi}\over 1 + \varepsilon \sin B 
\pi} + 2 \varepsilon \,, \quad 
c_{2} ={r^{2} e^{\varepsilon 2 \varphi}\over (1 + \varepsilon \sin B 
\pi)^{2}} \,.
\ee
Also,
\be 
\tr_{0} \R^{(\varepsilon)}_{0}( {i\pi\over 2} + \theta \,; \varphi)^{t_{0}}\ 
{\cal P}_{01}\ R_{01}(2 \theta)^{t_{1}} = g(\theta)\
\R^{(\varepsilon)}_{1}( {i\pi\over 2} - \theta \,; \varphi) \,,
\ee
where
\be
g(\theta) = {\varepsilon\sinh \theta - i \sin \pi B\over \sinh \theta} \,. 
\label{g}
\ee 
Setting
\be
\Y^{(\varepsilon)}(\theta \,; \varphi ) = 
\Y_{0}^{(\varepsilon)}(\theta)\ \Y_{1}^{(\varepsilon)}(\theta \,; \varphi )
\,,
\label{boundaryY}
\ee
it follows that $\Y_{0}^{(\varepsilon)}(\theta)$ and 
$\Y_{1}^{(\varepsilon)}(\theta \,; \varphi )$ must satisfy 
\be
\Y_{0}^{(\varepsilon)}(\theta)\
\Y_{0}^{(\varepsilon)}(-\theta)\
\cosh \theta  = 1 \,, \qquad 
\Y_{0}^{(\varepsilon)}({i\pi\over 2} + \theta)\ Y(2\theta)\ g(\theta)
= \Y_{0}^{(\varepsilon)}({i\pi\over 2} - \theta) \,,
\label{Y0properties}
\ee
and
\be
\Y_{1}^{(\varepsilon)}(\theta \,; \varphi )
\Y_{1}^{(\varepsilon)}(-\theta \,; \varphi )
\left( c_{0} + c_{1} \sinh^{2}{\theta\over 2} +
c_{2} \sinh^{4}{\theta\over 2} \right) = 1 \,, \quad  
\Y_{1}^{(\varepsilon)}({i\pi\over 2} + \theta \,; \varphi )
= \Y_{1}^{(\varepsilon)}({i\pi\over 2} - \theta \,; \varphi ) \,, 
\non \\ 
\label{Y1properties}
\ee
respectively.

For simplicity, we shall henceforth restrict 
our attention to the case $\varepsilon=+1$, and so 
we shall drop the superscript $(\varepsilon)$.
We propose the following integral representations for 
$\Y_{0}(\theta)$ and $\Y_{1}(\theta \,; \varphi)$:
\be
\Y_{0}(\theta) &=& {i\over \sqrt{2} \sinh({\theta\over 2} 
+{i \pi\over 4})} \exp \left( {1\over 2} \int_{0}^{\infty}
{dt\over t}\
{\sinh (2 i t \theta/\pi) \sinh(t(1+B)) \sinh (t B)\over \cosh^{2} t 
\cosh^{2}{t\over 2}} \right) \,, \non \\ 
\Y_{1}(\theta \,; \varphi) &=& {1\over r \cosh \varphi}
\exp \left( 4 \int_{0}^{\infty} {dt\over t}\
{\cosh (t \zeta/\pi) \cosh({t\over 2}(1-2B))  
\sinh ({t\over 2}(1 + {i \theta\over \pi}))
\sinh ({i t \theta\over 2 \pi}) \over \sinh t 
\cosh{t\over 2}} \right) \,, \non  \\
\label{boundaryYfactors}
\ee
where $\zeta$ is a function of the boundary parameter $\varphi$
defined by
\be
\zeta = \cos^{-1}\left(1 + e^{-2 \varphi}(1 + \sin B \pi) \right) \,.
\label{zeta}
\ee

In order to streamline the notation, let us denote the set of boundary 
parameters $\{ \eta \,, \vartheta \,, \varphi\}$ by $\xi$,
and denote the total scalar factor by $\Z(\theta \,; \xi)$
\be
\Z(\theta \,; \xi) &=& \SSS_{ShG}(\theta \,; \eta \,, \vartheta )\ 
\Y(\theta \,; \varphi ) \non \\
&=& \Z_{0}(\theta)\ \Z_{1}(\theta \,; \xi) \,.
\label{boundaryZ}
\ee
The proposed SShG boundary $S$ matrix is then given by
\be
\SSS(\theta \,; \xi) = \Z(\theta \,; \xi)\ 
\R(\theta \,; \varphi ) \,.
\label{boundaryS}
\ee 

We now observe that the boundary $S$ matrix is also invariant under
the strong-weak duality transformation (\ref{duality}).  Indeed, it is
evident that the matrix $\R(\theta \,; \varphi )$
(\ref{boundaryRmatrix}) has this invariance, if we assume that the
parameter $\varphi$ remains invariant under this transformation.  Let
us now consider the scalar factor.  The part of the scalar factor that
does not depend on boundary parameters can be written in the form
\be 
\Z_{0}(\theta) &=& \X_{0}(\theta)\ \Y_{0}(\theta) \\
&=& {i\over \sqrt{2} \sinh({\theta\over 2} 
-{i \pi\over 4})} \non \\
&\times& \exp \left( -{1\over 4} \int_{0}^{\infty}
{dt\over t}\
{\sinh (2 i t \theta/\pi) \over \cosh^{2} t \cosh^{2}{t\over 2}} 
\left[\cosh(t(1-2B)) (1 + 2 \cosh t) + \cosh t \right] \right) \,,
\label{boundaryZ0}
\ee
in which the duality invariance is manifest.  (The factors
$\X_{0}(\theta)$ and $\Y_{0}(\theta)$ are not separately invariant.) 
Finally, the part of the scalar factor which does depend on boundary
parameters,
\be
\Z_{1}(\theta \,; \xi) = \X_{1}(\theta \,; {4 \eta B\over \pi})\
\X_{1}(\theta \,; {4 i \vartheta B\over \pi})\  
\Y_{1}(\theta \,; \varphi)
\,,
\label{boundaryZ1}
\ee 
is also invariant under duality, since each
of its factors are separately invariant (provided the boundary
parameters $\eta B$ and $\vartheta B$ are assumed to remain invariant).

\section{Yang equations}

Having specified the bulk and boundary $S$ matrices, we are ready to 
start the TBA program. The first step is to formulate the Yang matrix 
and relate it to a commuting transfer matrix. Since this is not 
obvious for the case of boundaries, we begin by reviewing the more 
familiar case of periodic boundary conditions.

\subsection{Closed}

We consider $N$ particles of mass $m$ with real rapidities 
$\theta_{1} \,, \ldots \,, \theta_{N}$ and two-particle $S$ matrix 
$S(\theta)$ (\ref{bulkS}) in a periodic box of length $L >> {1\over 
m}$. The Yang equation \cite{Ya}, \cite{ZZ2} for particle 1 (which has 
momentum $P_{1} = m \sinh \theta_{1}$) is given by 
\be
\left( e^{i L m \sinh \theta_{1}} Y_{(1)} - \id \right) 
| \theta_{1} \,, \ldots \theta_{N} \rangle = 0 \,,
\label{closedYangEq}
\ee
where $Y_{(1)}$ is the ``Yang matrix''\footnote{We remind the reader 
that we are using the convention explained in Footnote \ref{convention}.}
\be
Y_{(1)} = S_{1 N}(\theta_{1} - \theta_{N})\ 
S_{1, N-1}(\theta_{1} - \theta_{N-1}) \cdots 
S_{1 2}(\theta_{1} - \theta_{2}) \,,
\label{closedYangMatrix}
\ee
which acts on $V^{\otimes N}$. There are similar equations, and 
corresponding matrices $Y_{(i)}$, for the other particles $i = 2 \,, 
3 \,, \ldots \,, N$.

\begin{figure}[htb]
	\centering
	\epsfxsize=0.4\textwidth\epsfbox{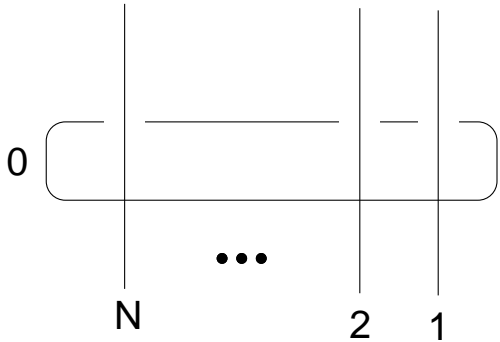}
	\caption[xxx]{\parbox[t]{.4\textwidth}{
	Closed-chain transfer matrix.}
	}
	\label{figclosed}
\end{figure}

The objective is to diagonalize $Y_{(i)}$.  The key to this problem is
to relate $Y_{(i)}$ to an inhomogeneous closed-chain transfer matrix,
for which there are well-developed diagonalization techniques.  (For
reviews, see e.g. \cite{KS} - \cite{Ne}.)  Indeed, consider the
transfer matrix (see Figure \ref{figclosed})
\be
\tau_{closed}(\theta | \theta_{1} \,, \ldots \,, \theta_{N})
= \tr_{0} \{ S_{0 N}(\theta - \theta_{N}) \cdots 
S_{0 2}(\theta - \theta_{2})\ S_{0 1}(\theta - \theta_{1}) \} \,,
\label{closedtransfer}
\ee
with inhomogeneities $\theta_{1} \,, \ldots \,, \theta_{N}$. Notice 
that we have introduced an additional (``auxiliary'') 2-dimensional 
vector space denoted by $0$. The product of $S$ matrices inside the 
trace (the so-called monodromy matrix) acts on $V^{\otimes (N+1)}$; 
but after performing the trace over the auxiliary space, one is left 
with an operator which acts on the (``quantum'') space $V^{\otimes 
N}$. Because $S(\theta)$ satisfies the Yang-Baxter equation, the 
transfer matrix commutes for different values of $\theta$
\be
\left[ \tau_{closed}(\theta | \theta_{1} \,, \ldots \,, \theta_{N}) \,,
\tau_{closed}(\theta' | \theta_{1} \,, \ldots \,, \theta_{N})  
\right] = 0 \,.
\label{closedcommutativity}
\ee

Let us now evaluate this transfer matrix at $\theta = \theta_{1}$. 
Using the fact that $S(0) = {\cal P}$ (the permutation matrix 
(\ref{permutation})) and ${\cal P}^{2} = \id $, we see that
\be
\tau_{closed}(\theta_{1} | \theta_{1} \,, \ldots \,, \theta_{N})
= \tr_{0} \{ ({\cal P}_{01} {\cal P}_{01}) S_{0 N}(\theta_{1} - \theta_{N}) 
\cdots ({\cal P}_{01} {\cal P}_{01}) 
S_{0 2}(\theta_{1} - \theta_{2})\ {\cal P}_{01} \} \,.
\ee
Finally, using ${\cal P}_{01}\ S_{0 i}\ {\cal P}_{01}= S_{1 i}$ and
$\tr_{0} {\cal P}_{01} = \id_{1} $, we conclude that
$\tau_{closed}(\theta_{1} | \theta_{1} \,, \ldots \,, \theta_{N}) = 
Y_{(1)}$. In general, we have
\be
Y_{(i)} = \tau_{closed}(\theta_{i} | \theta_{1} \,, \ldots \,, \theta_{N})
\,, \qquad i = 1 \,,  \ldots \,, N \,.
\label{closedresult}
\ee
This is the sought-after relation. In order to diagonalize the Yang 
matrices $Y_{(i)}$, it suffices to diagonalize the commuting 
closed-chain transfer matrix 
$\tau_{closed}(\theta | \theta_{1} \,, \ldots \,, \theta_{N})$.
That calculation, as well as the corresponding bulk TBA analysis, is 
described in \cite{Ah1}.

\subsection{Open}

\begin{figure}[htb]
	\centering
	\epsfxsize=0.4\textwidth\epsfbox{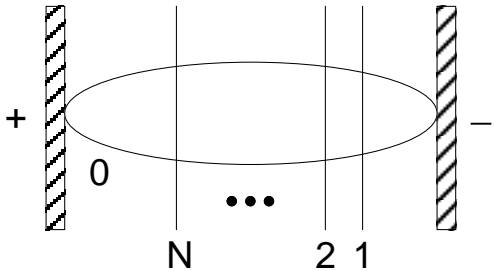}
	\caption[xxx]{\parbox[t]{.4\textwidth}{
	Open-chain transfer matrix.}
	}
	\label{figopen}
\end{figure}

We now turn to the case with boundaries, which is our primary interest
in this paper.  We therefore consider $N$ particles of mass $m$ with
real rapidities $\theta_{1} \,, \ldots \,, \theta_{N}$ in an interval
of length $L >> {1\over m}$, with bulk $S$ matrix $S(\theta)$
(\ref{bulkS}) and boundary $S$ matrix $\SSS(\theta \,; \xi)$
(\ref{boundaryS}).  The Yang equation for particle 1 is given by
\cite{FS}, \cite{GMN}
\be
\left( e^{2 i L m \sinh \theta_{1}} Y_{(1)} - \id \right) 
| \theta_{1} \,, \ldots \theta_{N} \rangle = 0 \,,
\label{openYangEq}
\ee
where the Yang matrix $Y_{(1)}$ is now given by
\be
Y_{(1)} = \SSS_{1}(\theta_{1}\,; \xi_{-})\ S_{2 1}(\theta_{1} + \theta_{2})
\cdots S_{N 1}(\theta_{1} + \theta_{N})\ 
\SSS_{1}(\theta_{1}\,; \xi_{+})\ S_{1 N}(\theta_{1} - \theta_{N})
\cdots S_{12}(\theta_{1} - \theta_{2}) \,,
\label{openYangMatrix}
\ee
where the subscripts $\pm$ denote the left and right boundaries. 
(There are similar matrices $Y_{(i)}$ for the other particles.) In 
analogy with the case of periodic boundary conditions, the key to 
diagonalizing the Yang matrix is to relate it to an inhomogeneous 
open-chain transfer matrix \cite{Sk} (see Figure \ref{figopen})
\be
\tau(\theta | \theta_{1} \,, \ldots \,, \theta_{N})
&=& \tr_{0} \{ \SSS_{0}(-\theta + i \pi\,; \xi_{+})^{t_{0}}\
S_{0 N}(\theta - \theta_{N}) \cdots S_{0 1}(\theta - \theta_{1}) 
\non \\
& \times &\SSS_{0}(\theta \,; \xi_{-})\
S_{0 1}(\theta + \theta_{1}) \cdots S_{0 N}(\theta + \theta_{N}) \} \,,
\label{opentransfer}
\ee
which commutes for different values of $\theta$
\be
\left[ \tau(\theta | \theta_{1} \,, \ldots \,, \theta_{N}) \,,
\tau(\theta' | \theta_{1} \,, \ldots \,, \theta_{N})  
\right] = 0 \,.
\label{opencommutativity}
\ee
Using the boundary cross-unitarity relation (\ref{boundarycrossing})
as well as the Yang-Baxter equation (\ref{YangBaxter}), 
(\ref{boundaryYangBaxter}), one can show that
\be
Y_{(i)} = \tau(\theta_{i} | \theta_{1} \,, \ldots \,, \theta_{N})
\,, \qquad i = 1 \,,  \ldots \,, N \,.
\label{openresult}
\ee
A proof for the case $N=2$ is presented in Appendix A. \footnote{We 
therefore fill a gap left open in \cite{FS}, where it was first 
observed that the open-chain Yang matrix is related to the Sklyanin 
transfer matrix; but neither the precise form of the relation nor its 
proof was given.} Hence, in order to diagonalize the Yang matrices
$Y_{(i)}$, it suffices to diagonalize the commuting open-chain 
transfer matrix $\tau(\theta | \theta_{1} \,, \ldots \,, \theta_{N})$. 
It is to this task that we devote the following section. 

\section{Inversion identity and transfer-matrix eigenvalues}

In this Section, we consider the problem of determining the 
eigenvalues of the inhomogeneous open-chain transfer matrix 
(\ref{opentransfer}). Our approach will be to first derive an exact 
so-called inversion identity. This approach has been used in the past 
to diagonalize simple (e.g., Ising) closed-chain transfer matrices 
\cite{Ba},\cite{Za2},\cite{Ah1}.

\subsection{Inversion identity}

Instead of working with the ``dressed'' transfer 
matrix (\ref{opentransfer}), it is more convenient (see Footnote 
\ref{period} below) to strip away the scalar factors from the bulk and 
boundary $S$ matrices, and to work instead with the ``bare'' transfer 
matrix
\be
\ttt(\theta | \theta_{1} \,, \ldots \,, \theta_{N}) &=& \tr_{0} \{ 
\R_{0}(-\theta + i \pi\,; \varphi_{+})^{t_{0}}\
R_{0 N}(\theta - \theta_{N}) \cdots R_{0 1}(\theta - \theta_{1}) \non \\
&\times& \R_{0}(\theta \,; \varphi_{-})\
R_{0 1}(\theta + \theta_{1}) \cdots R_{0 N}(\theta + \theta_{N}) \} \,,
\label{baretransfer}
\ee
where $R(\theta)$ is given by (\ref{bulkRmatrix}) and 
$\R(\theta\,; \varphi)$ is given by (\ref{boundaryRmatrix}) with 
$\varepsilon = +1$.

\begin{figure}[htb]
	\centering
	\epsfxsize=0.4\textwidth\epsfbox{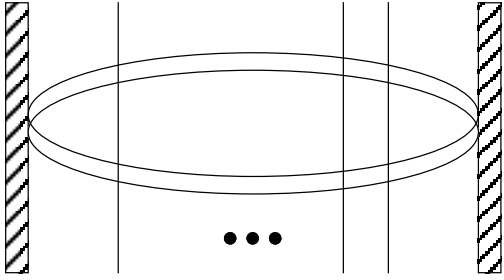}
	\caption[xxx]{\parbox[t]{.4\textwidth}{
	Fused open-chain transfer matrix.}
	}
	\label{figfused}
\end{figure}

There are two key points involved in obtaining the inversion identity. 
The first key point is to observe that the bulk $S$ matrix degenerates 
into a one-dimensional projector for a certain value of 
$\theta\ (= - i \pi)$:
\be
S(- i \pi) \propto {1\over 2}  \left( \begin{array}{rrrr}
  1  &0  &0  &-1  \\ 
  0  &0  &0  &0  \\
  0  &0  &0  &0 \\
 -1  &0  &0  &1 
\end{array} \right) \,.
\label{projector}
\ee 
Hence, it is possible to ``fuse'' \cite{Ka},\cite{KRS},\cite{KS} in
the auxiliary space, and thereby obtain a fusion formula of the form
\cite{MN}
\be
\ttt(\theta | \theta_{1} \,, \ldots \,, \theta_{N})\
\ttt(\theta + i \pi | \theta_{1} \,, \ldots \,, \theta_{N})
\propto \tilde \ttt(\theta | \theta_{1} \,, \ldots \,, \theta_{N}) + 
\Delta \,,
\label{roughfusionformula}
\ee
where $\tilde \ttt(\theta | \theta_{1} \,, \ldots \,, \theta_{N})$ is
a ``fused'' open-chain transfer matrix (see Figure \ref{figfused}), 
and here $\Delta$ represents a product of certain quantum determinants
\cite{IK}.  The fused transfer matrix is constructed from the fused
bulk $S$ matrix $\tilde R(\theta)$ and the fused boundary $S$ matrix
$\tilde \R(\theta \,; \varphi)$ , using the ``fused'' 3-dimensional
(instead of 2-dimensional) auxiliary space.

The second key point is that both $\tilde R(\theta)$ and 
$\tilde \R(\theta \,; \varphi)$ can be brought to upper-triangular 
form by a $\theta$-independent similarity transformation. This 
remarkable fact is presumably due to the fact that $R(\theta)$ 
satisfies the free Fermion condition (\ref{freeFermion}) (cf, 
\cite{Fe}). As a result, the fused transfer matrix is proportional to 
the identity matrix
\be
\tilde \ttt(\theta | \theta_{1} \,, \ldots \,, \theta_{N}) \propto \id
\,.
\ee
It follows from the fusion formula that the transfer matrix obeys an 
exact inversion identity
\be
\ttt(\theta | \theta_{1} \,, \ldots \,, \theta_{N})\
\ttt(\theta + i \pi | \theta_{1} \,, \ldots \,, \theta_{N})
= f(\theta) \id \,,
\label{inversionidentity1}
\ee
where $f(\theta)$ is a calculable scalar function. We find (see 
Appendix B for more details)
\be
\lefteqn{f(\theta) = {16 \sinh^{2} \theta\over 
\sinh(\theta - i B \pi) \sinh(\theta + i B \pi) 
(1 + \sin B \pi)^{2} \sin^{2} B \pi} 
\Big\{  }\non \\
& & \Big( \aaa \prod_{j=1}^{N} 
{\cosh({1\over 2}(\theta - \theta_{j})- i B \pi)\over
\cosh({1\over 2}(\theta - \theta_{j}))}
{\cosh({1\over 2}(\theta + \theta_{j})- i B \pi)\over
\cosh({1\over 2}(\theta + \theta_{j}))} \non \\ 
& &- \bbb \prod_{j=1}^{N} 
{\sinh({1\over 2}(\theta - \theta_{j})- i B \pi)\over
\sinh({1\over 2}(\theta - \theta_{j}))}
{\sinh({1\over 2}(\theta + \theta_{j})- i B \pi)\over
\sinh({1\over 2}(\theta + \theta_{j}))} \Big) \non \\
& &\times 
\Big( \ccc \prod_{j=1}^{N} 
{\cosh({1\over 2}(\theta - \theta_{j})+ i B \pi)\over
\cosh({1\over 2}(\theta - \theta_{j}))}
{\cosh({1\over 2}(\theta + \theta_{j})+ i B \pi)\over
\cosh({1\over 2}(\theta + \theta_{j}))} \non \\ 
& &- \ddd \prod_{j=1}^{N} 
{\sinh({1\over 2}(\theta - \theta_{j})+ i B \pi)\over
\sinh({1\over 2}(\theta - \theta_{j}))}
{\sinh({1\over 2}(\theta + \theta_{j})+ i B \pi)\over
\sinh({1\over 2}(\theta + \theta_{j}))} \Big)  \Big\} \,,
\label{functionf1}
\ee
where 
\be
\aaa &=& \sinh^{2}({1\over 2}(
i\pi B + {i\pi\over 2} - \theta))
\left[e^{-\varphi_{-}}\sinh^{2}({1\over 2}(i\pi B + {i\pi\over 2}))
+ e^{\varphi_{-}}\sinh^{2}({1\over 2}(
-i\pi B + {i\pi\over 2} + \theta)) \right] 
\times \left[\varphi_{-} \rightarrow \varphi_{+} \right] \non \\
\bbb &=& \sinh^{2}({1\over 2}(-i\pi B + {i\pi\over 2} + \theta))
\left[e^{-\varphi_{-}}\sinh^{2}({1\over 2}(i\pi B + {i\pi\over 2}))
+ e^{\varphi_{-}}\sinh^{2}({1\over 2}(
i\pi B + {i\pi\over 2} - \theta)) \right] 
\times \left[\varphi_{-} \rightarrow \varphi_{+} \right] \non \\
\ccc &=& \sinh^{2}({1\over 2}(i\pi B + {i\pi\over 2} + \theta))
\left[e^{-\varphi_{-}}\sinh^{2}({1\over 2}(i\pi B + {i\pi\over 2}))
+ e^{\varphi_{-}}\sinh^{2}({1\over 2}(
i\pi B - {i\pi\over 2} + \theta)) \right] 
\times \left[\varphi_{-} \rightarrow \varphi_{+} \right] \non \\
\ddd &=& \sinh^{2}({1\over 2}(i\pi B - {i\pi\over 2} + \theta))
\left[e^{-\varphi_{-}}\sinh^{2}({1\over 2}(i\pi B + {i\pi\over 2})
+ e^{\varphi_{-}}\sinh^{2}({1\over 2}(
i\pi B + {i\pi\over 2} + \theta)) \right] 
\times \left[\varphi_{-} \rightarrow \varphi_{+} \right] \,.\non  \\
\label{functionf2}
\ee
Notice that the function $f(\theta)$ is invariant under the duality
transformation $B \rightarrow 1 - B$. This inversion identity is one 
of the main results of this paper. We have checked it numerically up 
to $N=3$.

\subsection{Eigenvalues}

We now proceed to determine the eigenvalues of the transfer matrix. 
First, observe that by virtue of the commutativity property
(\ref{opencommutativity}), the bare transfer matrix 
$\ttt(\theta | \theta_{1} \,, \ldots \,, \theta_{N})$ has eigenstates
$| \theta_{1} \,, \ldots \theta_{N} \rangle$ which are independent 
of $\theta$,
\be
\ttt(\theta | \theta_{1} \,, \ldots \,, \theta_{N}) 
| \theta_{1} \,, \ldots \theta_{N} \rangle =
\LL (\theta | \theta_{1} \,, \ldots \,, \theta_{N})
| \theta_{1} \,, \ldots \theta_{N} \rangle \,,
\label{eigenvalueproblem}
\ee
where $\LL (\theta | \theta_{1} \,, \ldots \,, \theta_{N})$ are 
the corresponding eigenvalues.  Acting on $| \theta_{1} \,, \ldots 
\theta_{N} \rangle$ with the inversion identity, we obtain the 
corresponding identity for the eigenvalues
\be
\LL(\theta | \theta_{1} \,, \ldots \,, \theta_{N})\
\LL(\theta + i \pi | \theta_{1} \,, \ldots \,, \theta_{N})
= f(\theta) \,.
\label{inversionidentity2}
\ee

Moreover, one can show that the bare transfer matrix
$\ttt(\theta | \theta_{1} \,, \ldots \,, \theta_{N})$ is a periodic 
function of $\theta$ with period $2 \pi i$ \footnote{\label{period}
This is not the case for the dressed transfer matrix $\tau(\theta |
\theta_{1} \,, \ldots \,, \theta_{N})$, due to the presence of the
scalar factors.}
\be
\ttt(\theta + 2 \pi i| \theta_{1} \,, \ldots \,, \theta_{N}) = 
\ttt(\theta | \theta_{1} \,, \ldots \,, \theta_{N}) \,,
\label{periodicity}
\ee
whose asymptotic behavior for large $\theta$ is given by
\be
\ttt(\theta | \theta_{1} \,, \ldots \,, \theta_{N}) \sim 
{c\over 32} e^{3 \theta} \id  \quad \mbox{ for } 
\quad \theta \rightarrow \infty \,,
\label{asymptotic}
\ee
where
\be
c = {4 i e^{\varphi_{-} + \varphi_{+}}\over 
(1 + \sin B \pi) \sin B \pi} \,.
\label{cvalue}
\ee
Correspondingly, the eigenvalues obey
\be
\LL(\theta + 2 \pi i| \theta_{1} \,, \ldots \,, \theta_{N}) &=& 
\LL(\theta | \theta_{1} \,, \ldots \,, \theta_{N}) \,, \non  \\
\LL(\theta | \theta_{1} \,, \ldots \,, \theta_{N}) &\sim&
{c\over 32} e^{3 \theta} \quad \mbox{ for } 
\quad \theta \rightarrow \infty \,.
\label{properties}
\ee

The eigenvalues $\LL(\theta | \theta_{1} \,, \ldots \,, 
\theta_{N})$ are uniquely determined by the zeros and poles of 
$f(\theta)$, together with periodicity and asymptotic behavior.  
Indeed, observe that $f(\theta)$ is a product of two factors.  Let 
$\theta= z_{k}^{+} \,, z_{k}^{-}$ be zeros of the first, second 
factors, respectively.  Then $z_{k}^{+}$ obeys
\be
\lefteqn{\prod_{j=1}^{N}
{\tanh({1\over 2}(z_{k}^{+} - \theta_{j}) - i B \pi)\over
\tanh({1\over 2}(z_{k}^{+} - \theta_{j}))}
{\tanh({1\over 2}(z_{k}^{+} + \theta_{j}) - i B \pi)\over
\tanh({1\over 2}(z_{k}^{+} + \theta_{j}))} = 
{\sinh^{2}({1\over 2}(i\pi B + {i\pi\over 2} - z_{k}^{+}))
\over
\sinh^{2}({1\over 2}(-i\pi B + {i\pi\over 2} + z_{k}^{+}))}} 
\non \\
&\times&\left[{e^{-\varphi_{-}}\sinh^{2}({1\over 2}(
i\pi B + {i\pi\over 2}))
+ e^{\varphi_{-}}\sinh^{2}({1\over 2}(
-i\pi B + {i\pi\over 2} + z_{k}^{+}))\over
e^{-\varphi_{-}}\sinh^{2}({1\over 2}(i\pi B + {i\pi\over 2}))
+ e^{\varphi_{-}}\sinh^{2}({1\over 2}(
i\pi B + {i\pi\over 2} - z_{k}^{+}))}\right]  
\times \left[\varphi_{-} \rightarrow \varphi_{+} \right] \,,
\label{zplus}
\ee
and $z_{k}^{-}$ obeys
\be
\lefteqn{\prod_{j=1}^{N}
{\tanh({1\over 2}(z_{k}^{-} - \theta_{j}) + i B \pi)\over
\tanh({1\over 2}(z_{k}^{-} - \theta_{j}))}
{\tanh({1\over 2}(z_{k}^{-} + \theta_{j}) + i B \pi)\over
\tanh({1\over 2}(z_{k}^{-} + \theta_{j}))} = 
{\sinh^{2}({1\over 2}(i\pi B + {i\pi\over 2} + z_{k}^{-}))
\over
\sinh^{2}({1\over 2}(i\pi B - {i\pi\over 2} + z_{k}^{-}))}} 
\non \\
&\times&\left[{e^{-\varphi_{-}}\sinh^{2}({1\over 2}(
i\pi B + {i\pi\over 2}))
+ e^{\varphi_{-}}\sinh^{2}({1\over 2}(i\pi B - {i\pi\over 2} + 
z_{k}^{-}))\over
 e^{-\varphi_{-}}\sinh^{2}({1\over 2}(i\pi B + {i\pi\over 2}))
+ e^{\varphi_{-}}\sinh^{2}({1\over 2}(i\pi B + {i\pi\over 2} +
z_{k}^{-}))}\right]  
\times \left[\varphi_{-} \rightarrow \varphi_{+} \right] \,.
\label{zminus}
\ee
These are our ``magnonic'' Bethe Ansatz equations.

It follows \footnote{Indeed, let us denote the right-hand-side of 
Eq. (\ref{factored}) by $F(\theta)$. We observe that both $f(\theta)$
and $F(\theta)$ have the same periodicity (namely, $i \pi$, which is 
half the period of $\LL(\theta | \theta_{1} \,, \ldots \,, \theta_{N})$), 
the same zeros and poles in the 
strip $-{i\pi\over 2} < \theta < {i\pi\over 2}$, and the same 
asymptotic behavior. (The apparent poles of $f(\theta)$ at $\theta = 
\pm i B \pi$ are canceled by corresponding zeros.) Hence, the 
function $g(\theta) = F(\theta)/f(\theta)$ is regular everywhere in 
the complex $\theta$ plane, and thus must be constant by Liouville's 
theorem. By considering the limit $\theta \rightarrow \infty$, we see 
that this constant must be $1$.}
that $f(\theta)$ can be represented as 
\be
f(\theta)=-{c^{2}\over 16}\sinh^{2}\theta {\prod_{k=0}^{N} 
\sinh(\theta - z_{k}^{+})\sinh(\theta + z_{k}^{+})
\sinh(\theta - z_{k}^{-})\sinh(\theta + z_{k}^{-})\over
\prod_{k=1}^{N} 
\sinh^{2}(\theta - \theta_{k})\sinh^{2}(\theta + \theta_{k})} \,.
\label{factored}
\ee
It now follows by similar arguments that
\be
\LL(\theta | \theta_{1} \,, \ldots \,, \theta_{N})
= c \sinh \theta {\prod_{k=0}^{N} 
\sinh ({1\over 2}(\theta - z_{k}^{+}))
\sinh ({1\over 2}(\theta + z_{k}^{+}))
\sinh ({1\over 2}(\theta - z_{k}^{-}))
\sinh ({1\over 2}(\theta + z_{k}^{-})) \over
\prod_{k=1}^{N} 
\sinh ({1\over 2}(\theta - \theta_{k}))
\sinh ({1\over 2}(\theta + \theta_{k}))
\cosh ({1\over 2}(\theta - \theta_{k}))
\cosh ({1\over 2}(\theta + \theta_{k}))} \,.
\non  \\
\label{eigenvalues}
\ee
is the unique solution to the inversion identity 
(\ref{inversionidentity2}) with the properties (\ref{properties}).
Note that there are $N+1$ pairs of roots $z_{k}^{\pm}$, whereas in 
the case of periodic boundary conditions \cite{Ah1} there are only $N$. 
The appearance of the additional pair of roots $z_{0}^{\pm}$ is due to 
the fact that the boundary $S$ matrix $\R(\theta \,; \varphi)$ is not 
diagonal. The existence of these roots is essential for obtaining the correct 
asymptotic behavior; and it can be easily checked for the case $N=0$.

In summary, the eigenvalues of the bare transfer matrix 
(\ref{baretransfer}) are given by (\ref{eigenvalues}), where 
$z_{k}^{\pm}$ satisfy Eqs. (\ref{zplus}), (\ref{zminus}).

\subsection{Structure of Bethe Ansatz roots}

Before performing the thermodynamic ($N \rightarrow \infty$) limit 
(which is the subject of the next section), it is necessary to first 
understand the structure of the Bethe Ansatz roots. Following 
\cite{Ah1}, we observe that the Bethe Ansatz 
Eqs. (\ref{zplus}), (\ref{zminus}) have roots of the form
\be
z_{k}^{+}=\left\{ \begin{array}{c}
    x_{k} + i B \pi \\
    x_{k} + i B \pi + i\pi 
    \end{array} \right. \,, \qquad
z_{k}^{-}=\left\{ \begin{array}{c}
    x_{k} - i B \pi \\
    x_{k} - i B \pi - i\pi 
    \end{array} \right.  \,, 
\label{rootform}
\ee
where $x_{k}$ are real and satisfy
\be
\lefteqn{\prod_{j=1}^{N} \left[
{\tanh({1\over 2}(x_{k} - \theta_{j} - i B \pi))\over
 \tanh({1\over 2}(x_{k} - \theta_{j} + i B \pi))}
{\tanh({1\over 2}(x_{k} + \theta_{j} - i B \pi))\over
 \tanh({1\over 2}(x_{k} + \theta_{j} + i B \pi))} \right]
{\sinh^{2}({1\over 2}({i\pi\over 2} + x_{k})) \over
 \sinh^{2}({1\over 2}({i\pi\over 2} - x_{k}))}}  \non \\
&\times& 
\left[{e^{-\varphi_{-}}\sinh^{2}({1\over 2}(i\pi B + {i\pi\over 2}))
+ e^{\varphi_{-}}\sinh^{2}({1\over 2}({i\pi\over 2} - x_{k}))\over
e^{-\varphi_{-}}\sinh^{2}({1\over 2}(i\pi B + {i\pi\over 2}))
+ e^{\varphi_{-}}\sinh^{2}({1\over 2}({i\pi\over 2} + x_{k}))}\right]  
\times \left[\varphi_{-} \rightarrow \varphi_{+} \right] = 1 \,, \non  \\
& & \qquad k = 0 \,, 1 \,, \ldots \,, N \,. 
\label{realBAE}
\ee
Evidently, for each $x_{k}$, there are 4 possible combinations of 
roots $(z_{k}^{+} \,, z_{k}^{-})$. However, by considering the limit 
$B \rightarrow 0$, one can argue that only 2 of these combinations 
are allowed, which we denote by $\epsilon_{k}=+1$ and 
$\epsilon_{k}=-1$, respectively:
\be
\epsilon_{k} &=& +1: \qquad (z_{k}^{+} =  x_{k} + i B \pi  
\,, \quad z_{k}^{-} =  x_{k} - i B \pi - i\pi)  \,, \non \\
\epsilon_{k} &=& -1: \qquad (z_{k}^{+} =  x_{k} + i B \pi + i\pi 
\,, \quad z_{k}^{-} =  x_{k} - i B \pi )  \,.
\label{epsilons}
\ee

Hence, the eigenvalues are specified by $\{ x_{k} \,, \epsilon_{k} 
\}$, $k = 0 \,, \ldots \,, N$:
\be
\LL(\theta | \theta_{1} \,, \ldots \,, \theta_{N})_{\epsilon_{0} 
\cdots \epsilon_{N}}
= c \sinh \theta {\prod_{k=0}^{N} 
\lambda_{\epsilon_{k}}(\theta - x_{k})\
\lambda_{-\epsilon_{k}}(\theta + x_{k}) \over
\prod_{k=1}^{N} 
{1\over 4}\sinh (\theta - \theta_{k})\sinh (\theta + \theta_{k})} \,,
\label{eigenvalue2}
\ee
where 
\be
\lambda_{\epsilon}(\theta) = 
\sinh({1\over 2}(\theta - \epsilon i B \pi))
\cosh({1\over 2}(\theta + \epsilon i B \pi)) \,,
\ee
$\epsilon_{k} = \pm 1$, and $x_{k}$ satisfy (\ref{realBAE}).

To close this section, we observe that the ``dressed'' transfer matrix
(\ref{opentransfer}) is simply related to the ``bare'' transfer matrix
(\ref{baretransfer}) by
\be
\tau(\theta | \theta_{1} \,, \ldots \,, \theta_{N}) = 
\Z(-\theta + i\pi \,; \xi_{+}) \Z(\theta \,; \xi_{-}) 
\prod_{k=1}^{N} \left[ Z(\theta-\theta_{k}) Z(\theta+\theta_{k}) 
\right] 
\ttt(\theta | \theta_{1} \,, \ldots \,, \theta_{N}) \,,
\ee
where the scalar factors $Z(\theta)$ and $\Z(\theta \,; \xi)$ are 
introduced in Eqs. (\ref{bulkZ}), (\ref{boundaryZ}). Hence, the 
eigenvalues $\LLL (\theta | \theta_{1} \,, \ldots \,, \theta_{N})$ of
$\tau(\theta | \theta_{1} \,, \ldots \,, \theta_{N})$ are given by
\be
\LLL (\theta | \theta_{1} \,, \ldots \,, \theta_{N})_{\epsilon_{0} 
\cdots \epsilon_{N}}  &=&
{\Z(\theta \,; \xi_{+})\ \Z(\theta \,; \xi_{-})\ Z(-2\theta)\over 
g({i \pi\over 2} - \theta)}
\prod_{k=1}^{N} Z(\theta-\theta_{k}) Z(\theta+\theta_{k}) \non  \\
&\times&
\LL(\theta | \theta_{1} \,, \ldots \,, \theta_{N})_{\epsilon_{0} 
\cdots \epsilon_{N}} \,,
\label{dressedeigenvalue}
\ee
where 
$\LL(\theta | \theta_{1} \,, \ldots \,, \theta_{N})_{\epsilon_{0} 
\cdots \epsilon_{N}}$ is given by (\ref{eigenvalue2}). Here we have 
used the fact
\be
\Z(-\theta + i\pi \,; \xi) = {\Z(\theta\,; \xi) Z(-2\theta)\over 
g({i \pi\over 2} - \theta)} \,,
\ee
which follows from the bulk unitarity (\ref{ShGproperties}), 
(\ref{bulkYproperties}) and boundary cross-unitarity 
(\ref{boundShGproperties}), (\ref{Y0properties}), (\ref{Y1properties})
relations.

\section{Thermodynamic Bethe Ansatz analysis}

\begin{figure}[htb]
	\centering
	\epsfxsize=0.4\textwidth\epsfbox{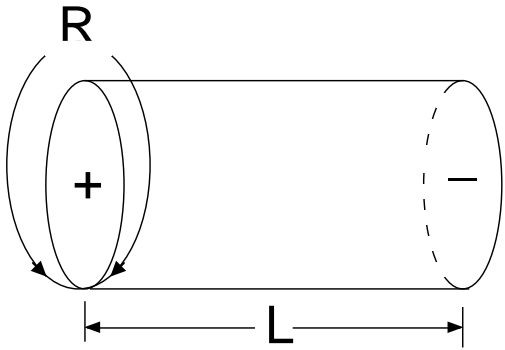}
	\caption[xxx]{\parbox[t]{.6\textwidth}{
	Cylinder on which the partition function $Z_{+-}$ is defined.}
	}
	\label{figcylinder}
\end{figure}

Having obtained the eigenvalues of the transfer matrix and the Bethe
Ansatz equations, we can proceed to the derivation of the TBA
equations and boundary entropy.  We begin by briefly reviewing the
general framework.  Following \cite{GZ},\cite{LMSS} we consider the
partition function $Z_{+-}$ of the system on a cylinder of length $L$
and circumference $R$ with left/right boundary conditions denoted by
$\pm$ (see Figure \ref{figcylinder})
\be
Z_{+-} &=& \tr e^{-R H_{+-}} = e^{- R F} \non \\
&=& \langle B_{+} | e^{- L H_{P}} | B_{-} \rangle \non \\
&\approx& \langle B_{+} | 0 \rangle  \langle 0 | B_{-} \rangle 
e^{- L E_{0}} \qquad \mbox{ for  } L \rightarrow \infty \,.
\ee
In the first line, Euclidean time evolves along the circumference of
the cylinder, and $H_{+-}$ is the Hamiltonian for the system with
spatial boundary conditions $\pm$.  In passing to the second line, we
rotate the picture, so that time evolves parallel to the axis of the
cylinder; $H_{P}$ is the Hamiltonian for the system with periodic
boundary conditions, and $| B_{\pm} \rangle $ are boundary states
which encode initial/final (temporal) conditions.  In the third line,
we consider the limit $L \rightarrow \infty$; the state $| 0 \rangle $
is the ground state of $H_{P}$, and $E_{0}$ is the corresponding
eigenvalue.  The quantity
$\ln \langle B_{+} | 0 \rangle \langle 0 | B_{-} \rangle$ is the
sought-after boundary entropy \cite{AL},\cite{LMSS}.  \footnote{More
precisely, we shall compute the dependence of the boundary entropy on
the boundary parameters.  The term in the boundary entropy which is
``constant'' (independent of boundary parameters) seems to be
difficult to compute even for simpler models \cite{LMSS},\cite{DRTW}.}
Taking the logarithm of the above expressions for the partition
function, one obtains
\be
- R F \approx - L E_{0} + 
\ln \langle B_{+} | 0 \rangle  \langle 0 | B_{-} \rangle
\,.
\label{be}
\ee
Whereas the free energy $F$ has a leading contribution which is of
order $L$, here we seek the subleading correction which is of order
$1$.

\subsection{Thermodynamic limit}

We proceed to compute $F$ using the TBA approach \cite{Ah1},
\cite{YY}-\cite{LMSS}.  To this end, we introduce the densities
$P_{\pm}(\theta)$ of ``magnons'', i.e., of real Bethe Ansatz roots 
$\{ x_{k} \}$ with $\epsilon_{k} = \pm 1$, respectively; and also the 
densities $\rho_{1}(\theta)$ and $\tilde\rho(\theta)$  of particles 
$\{ \theta_{k} \}$ and holes, respectively. Computing the logarithmic 
derivative of the ``magnonic'' Bethe Ansatz equations (\ref{realBAE}), we 
obtain \footnote{The term $-{1\over 2 \pi L}\Phi(\theta)$ originates 
from the exclusion \cite{FS},\cite{GMN} of the Bethe Ansatz root 
$x_{k}=0$.}
\be
P_{+}(\theta) + P_{-}(\theta) &=& {1\over 2\pi} \int_{0}^{\infty} 
d\theta'\ \rho_{1}(\theta') \left[ \Phi(\theta-\theta') + 
\Phi(\theta+\theta') \right] \non \\
&+& {1\over 2 \pi L}\left[-\Phi(\theta) + 
2 \Psi(\theta) + \Psi_{\varphi_{+}}(\theta) +
\Psi_{\varphi_{-}}(\theta) \right] \,,
\ee
where
\be 
\Phi(\theta) &=& {1\over i} {\partial\over \partial \theta} \ln \left(
{\tanh({1\over 2}(\theta - i B \pi))\over
 \tanh({1\over 2}(\theta + i B \pi))}\right) =
 {4 \cosh\theta \sin B \pi\over
 \cosh 2 \theta - \cos 2 B \pi} \,, \non  \\
\Psi(\theta) &=& {1\over i} {\partial\over \partial \theta} \ln \left(
{\sinh({1\over 2}({i \pi\over 2} + \theta)) \over
\sinh({1\over 2}({i \pi\over 2} - \theta))} \right) = - {1\over \cosh \theta} 
\,, \non  \\
\Psi_{\varphi}(\theta) &=& 
{1\over i} {\partial\over \partial \theta} \ln \left(
{e^{-\varphi}\sinh^{2}({1\over 2}(i\pi B + {i\pi\over 2}))
+ e^{\varphi}\sinh^{2}({1\over 2}({i\pi\over 2} - \theta))\over
e^{-\varphi}\sinh^{2}({1\over 2}(i\pi B + {i\pi\over 2}))
+ e^{\varphi}\sinh^{2}({1\over 2}({i\pi\over 2} + \theta))} \right)
\non  \\
&=& {4 \cosh\theta \cos \zeta \over \cosh 2 \theta + \cos 2 \zeta} \,,
\label{kernels}
\ee
where $\zeta$ is defined in (\ref{zeta}).
Defining $\rho_{1}(\theta)$ for negative values of $\theta$ to be 
equal to $\rho_{1}(|\theta|)$, we obtain the final form
\be
P_{+}(\theta) + P_{-}(\theta) ={1\over 2\pi} \left(
\rho_{1} * \Phi \right)(\theta) + {1\over 2 \pi L}\left[-\Phi(\theta) + 
2 \Psi(\theta) + \Psi_{\varphi_{+}}(\theta) +
\Psi_{\varphi_{-}}(\theta) \right] \,,
\label{constraint1}
\ee
where $*$ denotes convolution
\be
\left( f * g \right)(\theta) = \int_{-\infty}^{\infty} 
d\theta'\ f(\theta-\theta')g(\theta') \,.
\ee

We next consider the Yang equations, which imply 
(see Eqs. (\ref{openYangEq}),(\ref{openresult}))
\be
e^{2 i L m \sinh \theta_{k}} 
\LLL (\theta_{k} | \theta_{1} \,, \ldots \,, \theta_{N}) = 1 
\,, \qquad k = 1 \,, \ldots \,, N \,,
\ee
where $\LLL (\theta | \theta_{1} \,, \ldots \,, \theta_{N})$ 
is the eigenvalue of the dressed transfer matrix 
$\tau(\theta | \theta_{1} \,, \ldots \,, \theta_{N})$, 
which is given by (\ref{dressedeigenvalue}). Computing the 
logarithmic derivative, we obtain
\be
\rho_{1}(\theta) + \tilde\rho(\theta) &=& {1\over 2 \pi} \Big\{
2 m \cosh \theta + \int_{0}^{\infty}d\theta'\ \rho_{1}(\theta') 
\left[\Phi_{Z}(\theta-\theta') + \Phi_{Z}(\theta+\theta') \right] 
\non \\
&+& \int_{0}^{\infty}d\theta'\ \Big[ 
P_{+}(\theta')\Phi_{+}(\theta-\theta') +
P_{-}(\theta')\Phi_{-}(\theta-\theta') \non  \\ 
&+&
P_{-}(\theta')\Phi_{+}(\theta+\theta') +
P_{+}(\theta')\Phi_{-}(\theta+\theta') \Big] \non \\
&+& {1\over L}\left[-\Phi_{Z}(\theta) - 2\Phi_{Z}(2\theta) 
+ {\partial\over \partial \theta} Im \ln \Z(\theta \,; \xi_{+})
+ {\partial\over \partial \theta} Im \ln \Z(\theta \,; \xi_{-})
\right] \Big\} \,,
\ee
where
\be
\Phi_{Z}(\theta) = {\partial\over \partial \theta} Im \ln Z(\theta) 
\,, \qquad 
\Phi_{\pm}(\theta) = {\partial\over \partial \theta} Im \ln 
\lambda_{\pm}(\theta) \,.
\ee
Using the fact $\Phi_{\pm}(\theta) = \pm {1\over 2}\Phi(\theta)$, and 
defining $P_{\pm}(\theta)$ for negative values of $\theta$ to be 
equal to $-P_{\pm}(|\theta|)$, we obtain
\be
\rho_{1}(\theta) + \tilde\rho(\theta) &=& 
{m\over \pi} \cosh \theta 
+ {1\over 2 \pi} \left( \rho_{1} * \Phi_{Z}\right) (\theta) 
+ {1\over 4 \pi} \left( (P_{+} - P_{-}) * \Phi \right) (\theta)
\non \\
&+& {1\over 2 \pi L}\left[-\Phi_{Z}(\theta) - 2\Phi_{Z}(2\theta) 
+ {\partial\over \partial \theta} Im \ln \Z(\theta \,; \xi_{+})
+ {\partial\over \partial \theta} Im \ln \Z(\theta \,; \xi_{-})
\right] \,.
\ee
We now use (\ref{constraint1}) to eliminate $P_{-}$, and use 
the expressions (\ref{boundaryZ}), (\ref{boundaryZ1})
to separate the various factors in $\Z(\theta \,; \xi)$ to obtain
\be
\rho_{1}(\theta) + \tilde\rho(\theta) &=& 
{m\over \pi} \cosh \theta 
+ {1\over 2\pi}  P_{+} * \Phi 
+ {1\over 2 \pi} \rho_{1} * \left(\Phi_{Z} 
- {1\over 4 \pi} \Phi * \Phi \right) \non \\
&+& {1\over 2 \pi L}\Bigg[ 
-\left(\Phi_{Z} - {1\over 4 \pi} \Phi * \Phi \right)
+ 2 \left( {\partial\over \partial \theta} Im \ln \Z_{0}(\theta)
- \Phi_{Z}(2\theta) - {1\over 4 \pi} \Psi * \Phi \right) 
\non \\
&+& \left( {\partial\over \partial \theta} 
Im \ln \Y_{1}(\theta \,; \varphi_{+})
- {1\over 4 \pi} \Psi_{\varphi_{+}} * \Phi \right) 
+ \left( {\partial\over \partial \theta} 
Im \ln \Y_{1}(\theta \,; \varphi_{-})
- {1\over 4 \pi} \Psi_{\varphi_{-}} * \Phi \right) \non \\
&+& {1\over i}{\partial\over \partial \theta} 
\ln \X_{1}(\theta \,; {4 \eta_{+} B\over \pi})\
\X_{1}(\theta \,; {4 i \vartheta_{+} B\over \pi})
+{1\over i}{\partial\over \partial \theta} 
\ln \X_{1}(\theta \,; {4 \eta_{-} B\over \pi})\
\X_{1}(\theta \,; {4 i \vartheta_{-} B\over \pi})
\Bigg] \,. \non \\
\ee
Noting the ``bulk'' identity \cite{AR}
\be
\Phi_{Z}(\theta)  
- {1\over 4 \pi} \left( \Phi * \Phi \right)(\theta)  = 0 \,,
\ee
and its boundary counterparts
\be
{\partial\over \partial \theta} 
Im \ln \Y_{1}(\theta \,; \varphi)
- {1\over 4 \pi} \left( \Psi_{\varphi} * \Phi \right)(\theta) 
&=& 0 \,, \non  \\
{\partial\over \partial \theta} Im \ln \Z_{0}(\theta)
- \Phi_{Z}(2\theta) 
- {1\over 4 \pi} \left( \Psi * \Phi \right)(\theta)
&=& -{1\over 4}\Phi(\theta) + \Psi(\theta) \,, 
\ee 
we remain with the rather simple result
\be
\rho_{1}(\theta) + \tilde\rho(\theta) &=& 
{m\over \pi} \cosh \theta 
+ {1\over 2\pi}  \left( P_{+} * \Phi \right)(\theta)
+ {1\over 2 \pi L}\Bigg[ 
-{1\over 2}\Phi(\theta) + 2 \Psi(\theta) \non \\
&+& \kappa(\theta \,; {4 \eta_{+} B\over \pi})
+ \kappa(\theta \,; {4 i \vartheta_{+} B\over \pi})
+ \kappa(\theta \,; {4 \eta_{-} B\over \pi})
+ \kappa(\theta \,; {4 i \vartheta_{-} B\over \pi})
\Bigg] \,,
\label{constraint2}
\ee
where 
\be
\kappa(\theta \,; F) = {1\over i}{\partial\over \partial \theta} 
\ln \X_{1}(\theta \,; F) = {4 \cosh \theta \cos(\pi F/2)\over
\cosh 2\theta + \cos \pi F} \,.
\label{kappa}
\ee 

The thermodynamic limit of the magnonic Bethe Ansatz equations and the
Yang equations, given by (\ref{constraint1}) and (\ref{constraint2}),
respectively, are the main results of this subsection.  Notice that
the former depends on the boundary parameters $\varphi_{\pm}$, while
the latter depends on the (boundary sinh-Gordon) boundary parameters
$\eta_{\pm} \,, \vartheta_{\pm}$.

\subsection{TBA equations and boundary entropy}

The free energy $F$ is given by
\be
F = E - T S \,,
\ee
where the temperature is $T={1\over R}$, the energy $E$ is
\be
E = \sum_{k=1}^{N} m \cosh \theta_{k} 
= {L\over 2}\int_{-\infty}^{\infty}d\theta\ 
\rho_{1}(\theta) m \cosh \theta \,,
\ee
and the entropy $S$ is \cite{YY},\cite{Za2}
\be
S &=& {L\over 2}\int_{-\infty}^{\infty}d\theta\ \Big\{
(\rho_{1} + \tilde \rho) \ln (\rho_{1} + \tilde \rho) 
- \rho_{1}  \ln \rho_{1} - \tilde \rho \ln \tilde \rho \non \\
&+& (P_{+} + P_{-}) \ln (P_{+} +  P_{-}) 
- P_{+}  \ln P_{+} -  P_{-} \ln P_{-} \Big\} \,.
\ee
Extremizing the free energy $(\delta F = 0)$ subject to the constraints
\be
\delta P_{-} &=& -\delta P_{+} 
+ {1\over 2 \pi} \delta \rho_{1} * \Phi \,, \non  \\
\delta \tilde\rho  &=& -\delta \rho_{1} 
+ {1\over 2 \pi} \delta P_{+} * \Phi \,,
\ee
(which follow from Eqs.  (\ref{constraint1}), (\ref{constraint2}),
respectively) we obtain a set of TBA equations which is the same as
for the case of periodic boundary conditions \cite{Ah1},\cite{AR}
\be
r \cosh \theta &=& \epsilon_{1}(\theta) 
+ {1\over 2 \pi} \left( \Phi * L_{2} \right)(\theta) \,, \non  \\
0 &=& \epsilon_{2}(\theta) 
+ {1\over 2 \pi} \left( \Phi * L_{1} \right)(\theta) \,, 
\ee
where 
\be
L_{i}(\theta) &=& \ln \left( 1 + e^{-\epsilon_{i}(\theta)} \right) \,,
\qquad r = m R \,, \non \\
\epsilon_{1} &=& \ln \left( {\tilde \rho\over \rho_{1}} \right) \,, 
\qquad 
\epsilon_{2} = \ln \left( {P_{-}\over P_{+}} \right) \,.
\ee

We next evaluate $F$ using also the constraints (\ref{constraint1}),
(\ref{constraint2}) and the TBA equations.  From the boundary (order
$1$) contribution, we obtain (see Eq.  (\ref{be})) the boundary
entropy
\be
\ln \langle B_{+} | 0 \rangle  \langle 0 | B_{-} \rangle =
{1\over 4\pi}\int_{-\infty}^{\infty}d\theta\ &\Big\{ &
\Big[-{1\over 2}\Phi(\theta) + 2 \Psi(\theta) + 
\kappa(\theta \,; {4 \eta_{+} B\over \pi})
+ \kappa(\theta \,; {4 i \vartheta_{+} B\over \pi}) \non  \\
&+& \kappa(\theta \,; {4 \eta_{-} B\over \pi})
+ \kappa(\theta \,; {4 i \vartheta_{-} B\over \pi}) \Big] 
L_{1}(\theta) \non  \\
&+& \left[-\Phi(\theta) + 2 \Psi(\theta) + 
\Psi_{\varphi_{+}}(\theta) + \Psi_{\varphi_{-}}(\theta)
\right] L_{2}(\theta)
\,.
\ee
In particular, the dependence of the boundary entropy of a single 
boundary on the boundary parameters is given by \footnote{For the 
case $\varepsilon = -1$, we obtain a similar result, except the 
parameter $\zeta$ appearing in the kernel $\Psi_{\varphi}(\theta)$ is 
now given by 
$\zeta = \cos^{-1}\left(-1 + e^{2 \varphi}(1 - \sin B \pi) \right)$
instead of by Eq. (\ref{zeta}).}
\be
s_{B}(\eta \,, \vartheta \,, \varphi) =
{1\over 4\pi}\int_{-\infty}^{\infty}d\theta\ \left\{ 
\left[ \kappa(\theta \,; {4 \eta B\over \pi})
+ \kappa(\theta \,; {4 i \vartheta B\over \pi}) \right]
L_{1}(\theta)  
+ \Psi_{\varphi}(\theta)\ L_{2}(\theta) \right\}
\,,
\label{beresult}
\ee
where the kernels $\kappa(\theta \,; F)$ and $\Psi_{\varphi}(\theta)$ 
are defined in Eqs. (\ref{kappa}) and (\ref{kernels}), respectively.
The term involving $L_{1}$, which had previously been conjectured 
\cite{AR}, depends on the boundary sinh-Gordon parameters $\eta \,, 
\vartheta$. The term involving $L_{2}$, which had not been anticipated, 
depends on the boundary parameter $\varphi$ (which appears in 
$\R(\theta \,; \varphi)$, i.e., the non-diagonal part of the boundary 
$S$ matrix). This expression for the boundary entropy is another of 
the main results of this paper.

\section{Boundary roaming trajectories}

One application of our result (\ref{beresult}) for the boundary
entropy is to obtain boundary roaming trajectories corresponding to 
$c < 3/2$ superconformal models.  In order to best explain this result,
it is helpful to first recall earlier work on bulk and boundary
roaming.

Al. Zamolodchikov \cite{Za4} first considered the TBA equations for 
the bulk ShG (non-supersymmetric) model with the coupling constant 
$\gamma$ analytically continued to complex values,
\be
\gamma = {\pi\over 2} \pm i \theta_{0} \,, \qquad \theta_{0} >> 1 \,.
\ee
The corresponding effective central charge $c_{eff}(r)$ 
interpolates (``roams'') between the values
\be
c_{p} = 1 - {6\over p(p+1)} \,, \qquad p = 3 \,, 4\,, 5 \,, \ldots
\ee
corresponding to the unitary $c<1$ minimal models \cite{BPZ}. Indeed, 
a plot of $c_{eff}(r)$ vs. $\log(r/2)$ reveals a ``staircase'' with 
plateaus at values of $c_{eff}(r)$ equal to $c_{p}$.

This result was later generalized \cite{LSS} to the boundary ShG 
model: choosing the value of $r$ so that $c_{eff}(r)$ lies on some 
plateau, the boundary entropy $s_{B}(F)$ (where $F$ is a boundary 
parameter) interpolates between values corresponding to various 
conformal boundary conditions \cite{Ca}.

The original work \cite{Za4} was also generalized \cite{AR} to the 
bulk SShG (supersymmetric) model. The TBA equations with a similar 
analytic continuation of the coupling constant
\be
\pi B = {\pi\over 2} \pm i \theta_{0} \,, \qquad \theta_{0} >> 1 
\label{roaminglimit}
\ee
cause the effective central charge $c_{eff}(r)$ to
interpolate between the values
\be
c_{p} = {3\over 2}\left(1 - {8\over p(p+2)} \right) \,, 
\qquad p = 4 \,, 6\,, 8 \,, \ldots
\label{superconf}
\ee
corresponding to the even unitary $c<3/2$ minimal models \cite{BKT},
\cite{FQS}.  Precisely this set of TBA equations had been conjectured
earlier in \cite{Ma}, and then further generalized in \cite{DR}.

Finally, let us consider the model of primary interest here, namely, 
boundary SShG. 
For simplicity, we fix $\phi_0=0$ in the boundary Lagrangian 
(\ref{boundaryLagrangian}), which corresponds to $\vartheta=0$.\footnote{
Consider the boundary SSG model first. 
When $\phi_0=0$ the total Lagrangian respects $\sf C$ symmetry due to
the $Z_2$ symmetry $\phi\to -\phi$. Therefore, the boundary $S$ matrix
should respect $\sf C$ symmetry, namely the soliton and antisoliton 
should scatter equally on the boundary.  
Since the topological sector of the SSG $S$ matrix is encoded in the SG part,
the boundary parameter $\vartheta$ should vanish as it does in the SG 
model \cite{GZ}. 
This holds also for the boundary SShG $S$ matrix because
the two models are related by the fusion procedure.}
Due to the roaming limit (\ref{roaminglimit}), we should rescale the remaining 
two parameters $\eta,\varphi$ so that the boundary entropy can be a 
function of well-defined (finite) boundary parameters.
For this purpose we set $\eta\to 0$ and $\varphi\to\infty$ while keeping
$\theta_0\eta$ and $\theta_0-2\varphi$ finite. Let us introduce
new boundary parameters $f_1$ and $f_2$ defined by (see Eq. 
(\ref{zeta}))
\be
{2\theta_0\eta\over{\pi}}\equiv f_1,\qquad
1+{1\over{2}}e^{\theta_0-2\varphi}\equiv \cosh f_2 \,.
\ee
We can reexpress the boundary entropy (\ref{beresult}) in terms of 
these parameters as 
\be
s_B=s_{B}^{(1)}+s_{B}^{(2)}\,, \qquad \mbox{ where }\qquad 
s_{B}^{(i)}={1\over{4\pi}}\int_{-\infty}^{\infty}d\theta\ 
\Psi(\theta;f_i)\ L_i(\theta),\qquad i=1,2,
\ee
with
\be
\Psi(\theta;f)={4\cosh\theta\cosh f\over{\cosh 2\theta+\cosh 2f}}.
\ee
To compute the roaming boundary entropy, we fix a value of $r$ 
where $c_{eff}(r)$ lies on a plateau (\ref{superconf}). 
Then, as we change the boundary roaming parameters $f_1$ and $f_2$, we
check if the boundary entropy 
interpolates between the values \cite{AR} \footnote{Note that this expression 
satisfies $s_{B}(1 \,, 1)=0$. The correct expression for the conformal 
boundary entropies has an additional ``constant'' term (i.e., 
independent of both $r$ and $s$); we neglect this term here, since we 
are mostly interested in differences $s_{B}(r \,, s) - s_{B}(r' \,, 
s')$, for which the constant term cancels.}
\be
s_{B}(r \,, s) &=& \ln \left[ \left(
{\sin({\pi r \over p})\over \sin({\pi \over p})} \right)
\left(
{\sin({\pi s \over p+2})\over \sin({\pi \over p+2})} \right) \right] 
\non \\
 &=& s_{B}(r \,, 1) + s_{B}(1 \,, s) \,.
\label{additivity}
\ee
corresponding to conformal boundary states $(r \,, s)$ (which, in 
turn, correspond \footnote{We 
recall \cite{Ca} that for each bulk primary field $\Phi_{(r \,, s)}$, there 
corresponds a conformal boundary state $| \tilde h_{(r \,, s)} \rangle$
(which, for brevity, we denote here by $(r \,, s)$ ) such that the 
partition function $Z_{(1 \,, 1) (r \,, s)}$ for the CFT on a cylinder 
with conformal boundary states $(1 \,, 1)$ and $(r \,, s)$ is 
given by
\be
Z_{(1 \,, 1) (r \,, s)} = \chi_{(r \,, s)}(q) \,, \non 
\ee
i.e., the character of $\Phi_{(r \,, s)}$. In particular, 
\be
Z_{(1 \,, 1) (1 \,, 1)} = \chi_{(1 \,, 1)}(q)  \non 
\ee
is the character of the unit operator.}
to primary fields $\Phi_{(r \,, s)}$). 

\begin{figure}[htb]
	\centering
	\epsfxsize=0.6\textwidth\epsfbox{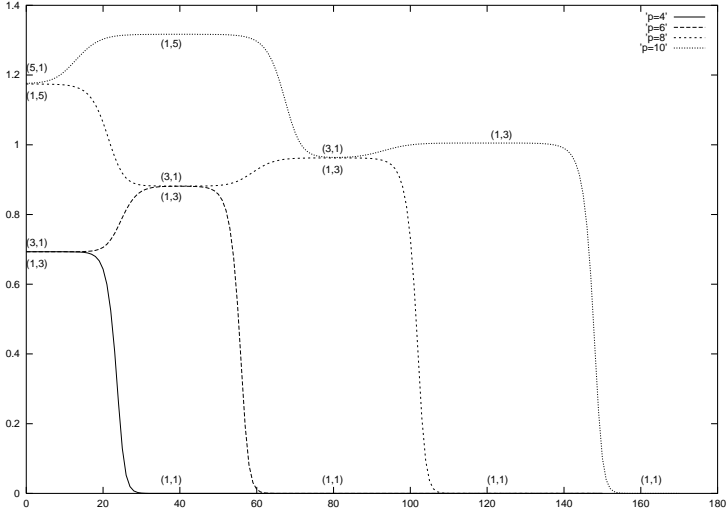}
	\caption[xxx]{\parbox[t]{0.5\textwidth}{
	Boundary roaming trajectories: $s_B^{(1)}$ vs. $f_1$.}
	}
	\label{figL1}
\end{figure}

\begin{figure}[htb]
	\centering
	\epsfxsize=0.6\textwidth\epsfbox{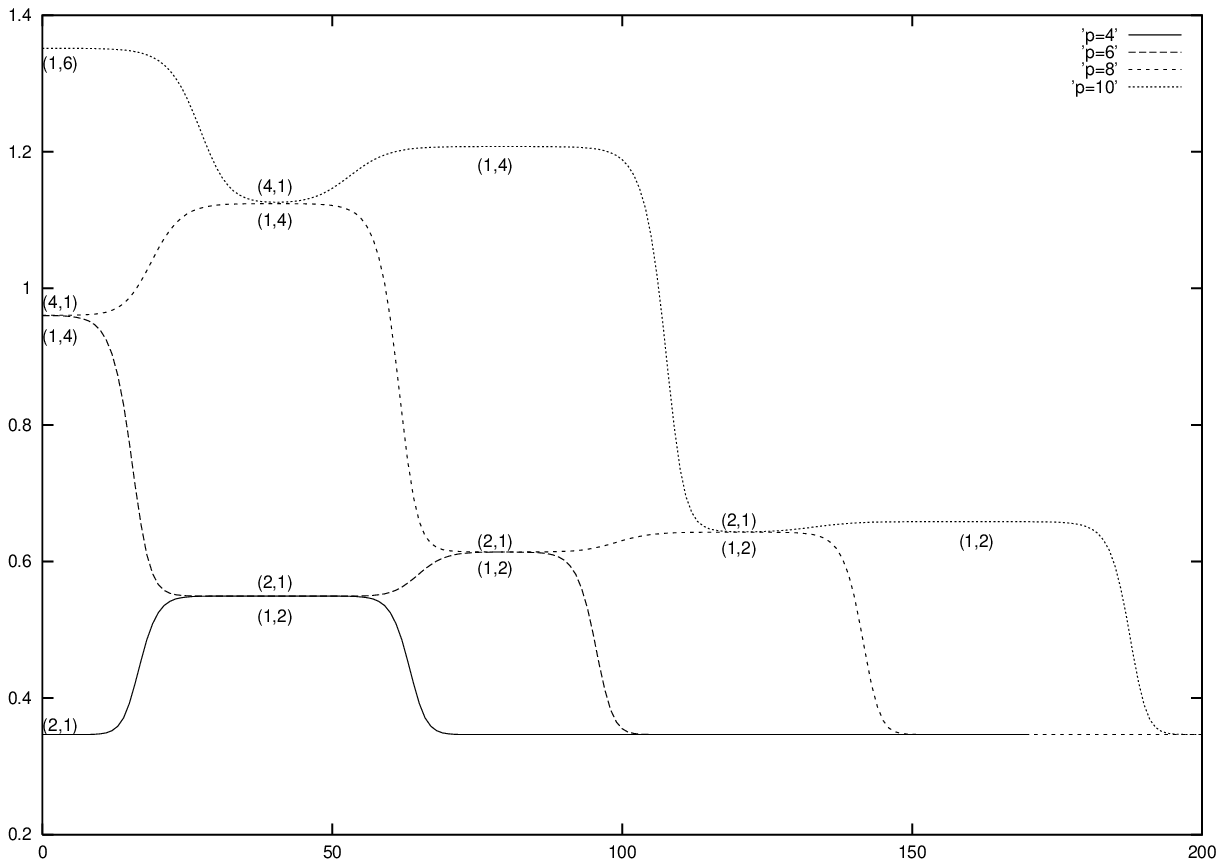}
	\caption[xxx]{\parbox[t]{0.5\textwidth}{
	Boundary roaming trajectories: $s_B^{(2)}$ vs. $f_2$.}
	}
	\label{figL2}
\end{figure}

Indeed, we can see clearly from Figure \ref{figL1} \cite{AR} that $s_{B}^{(1)}$
interpolates between boundary entropies of the conformal boundary states 
\be
(1\,, a) \leftrightarrow \left\{ \begin{array}{c}
    (a-2 \,, 1) \\
    (a \,, 1)
    \end{array} \right.  \,, \qquad a \mbox{  odd  } \,.
\ee
Similarly $s_{B}^{(2)}$ generates the new flow
(see Figure \ref{figL2}) \footnote{For $p>4$, we cannot associate 
any conformal boundary state to the final plateau (i.e., for
asymptotically large values of the boundary parameter $f_{2}$), since there 
is no state $(0 \,, 1)$.}
\be
(1\,, a) \leftrightarrow \left\{ \begin{array}{c}
    (a-2 \,, 1) \\
    (a \,, 1)
    \end{array} \right.  \,, \qquad a \mbox{  even  } \,.
\ee
While these flows are generated by changing one parameter while fixing 
the other, we can generate more general flows by changing $f_1$ and $f_2$ 
simultaneously.
In view of the additivity property (\ref{additivity}),
these two sets of flows can be combined to generate additional flows for
the total boundary entropy $s_B$
\be
(r\,, s) \leftrightarrow \left\{ \begin{array}{c}
    (s \,, r) \\
    (s-2 \,, r) \\
    (s \,, r+2) \\
    (s-2 \,, r+2) 
    \end{array} \right.  \,, \qquad r-s = \mbox{  odd  } \,.
\ee
Note that $r-s=$ even/odd corresponds to the Neveu-Schwarz/Ramond 
sectors, respectively.

\section{Discussion}

We have presented the exact solution of the boundary SShG model -- an
integrable QFT whose bulk and boundary $S$ matrices are not diagonal. 
In particular, we have derived an exact inversion identity
(\ref{inversionidentity1}) - (\ref{functionf2}), as well as the TBA
equations and boundary entropy (\ref{beresult}).  Moreover, we have
uncovered a rich pattern of boundary roaming trajectories, which
remain to be understood in detail.

Although the boundary SShG model has a special feature which allows it 
to be solved by an inversion identity (namely, the bulk $S$ matrix 
satisfies by free-Fermion condition (\ref{freeFermion})), it is by no 
means the only such model.  Indeed, there are infinite families of 
integrable QFTs with $N=1$ or $N=2$ supersymmetry 
\cite{Sc}-\cite{FI} that have this property.  These models have bulk 
and boundary $S$ matrices which are similar to those of SShG, and 
therefore, we expect similar inversion identities to hold.  We hope to 
report on these models in the near future \cite{AN}.

Finally, we recall \cite{Sk} that one can readily obtain the
Hamiltonian of an integrable open quantum spin chain with $N$ spins
from any homogeneous open-chain transfer matrix $\ttt(\theta | 0)$
(\ref{baretransfer}).  Indeed, the Hamiltonian ${\cal H}$ is given by
\be
{\cal H} \propto {\partial\over \partial \theta} \ttt(\theta | 0) 
\Big\vert_{\theta=0} \,,
\ee 
which commutes with $\ttt(\theta | 0)$.  For the $R$ matrices which we
have considered here (\ref{bulkRmatrix}), (\ref{boundaryRmatrix}), the
corresponding Hamiltonian is that of a certain anisotropic XY chain
with both bulk and boundary magnetic fields.  By determining the
eigenvalues (\ref{eigenvalues}) of the transfer matrix, we have
evidently also solved the corresponding open quantum spin chain.  It
would be interesting to exploit this solution to determine properties
of this model in the thermodynamic limit.

\section*{Acknowledgments}

We thank O. Alvarez, D. Bernard, E. Corrigan, G. Delius, P. Dorey, P.
Fendley, M. Martins and H. Saleur for helpful comments and/or
correspondence.  One of us (R.N.) is grateful for the hospitality at
the APCTP in Seoul (where this work was initiated) and at the CRM in
Montreal (where the results were first reported).  This work was
supported in part by KOSEF 1999-2-112-001-5 (C.A.) and by the National
Science Foundation under Grant PHY-9870101 (R.N.).

\appendix

\section{Relation of Yang matrix to Sklyanin transfer matrix}

In Section 3.2, we stated that the Yang matrix (\ref{openYangMatrix})
is related to the Sklyanin open-chain transfer matrix
(\ref{opentransfer}) in the following way (\ref{openresult}):
\be
Y_{(i)} = \tau(\theta_{i} | \theta_{1} \,, \ldots \,, \theta_{N})
\,, \qquad i = 1 \,,  \ldots \,, N \,.
\ee
We present here a proof for the case $N=2$. Evaluating the transfer 
matrix at $\theta=\theta_{1}$, we have
\be
\tau(\theta_{1} | \theta_{1} \,, \theta_{2})
&=& \tr_{0} \{ \SSS_{0}(-\theta_{1} + i \pi\,; \xi_{+})^{t_{0}}
S_{0 2}(\theta_{1} - \theta_{2}) S_{0 1}(0) 
\SSS_{0}(\theta_{1} \,; \xi_{-})
S_{0 1}(2\theta_{1}) S_{0 2}(\theta_{1} + \theta_{2}) \} \non \\
&=& \tr_{0} \{ S_{0 2}(\theta_{1} - \theta_{2}) {\cal P}_{0 1} 
\SSS_{0}(\theta_{1} \,; \xi_{-}) ({\cal P}_{0 1} {\cal P}_{0 1})
S_{0 1}(2\theta_{1}) ({\cal P}_{0 1} {\cal P}_{0 1}) \non \\
&\times& S_{0 2}(\theta_{1} + \theta_{2}) ({\cal P}_{0 1} {\cal P}_{0 1})
\SSS_{0}(-\theta_{1} + i \pi\,; \xi_{+})^{t_{0}}\} = \cdots 
\ee
In passing to the second line, we have used the cyclic property of 
the trace, as well as $S(0) = {\cal P}$ and ${\cal P}^{2} = \id $,
where ${\cal P}$ is the permutation matrix (\ref{permutation}).
\be
\cdots = \SSS_{1}(\theta_{1} \,; \xi_{-})
\tr_{0} \{ S_{0 2}(\theta_{1} - \theta_{2}) 
S_{0 1}(2\theta_{1}) S_{1 2}(\theta_{1} + \theta_{2})
{\cal P}_{0 1} \SSS_{0}(-\theta_{1} + i \pi\,; \xi_{+})^{t_{0}}\} 
= \cdots 
\ee
Here we have used ${\cal P}_{01}\ X_{0}\ {\cal P}_{01}= X_{1}$, and 
the $\sf P$ symmetry of the $R$ matrix (\ref{Rsymmetries}). 
\be
\cdots = \SSS_{1}(\theta_{1} \,; \xi_{-})
\tr_{0} \{ S_{1 2}(\theta_{1} + \theta_{2}) 
S_{0 1}(2\theta_{1}) S_{0 2}(\theta_{1} - \theta_{2}) 
{\cal P}_{0 1} \SSS_{0}(-\theta_{1} + i \pi\,; \xi_{+})^{t_{0}}\} 
= \cdots 
\ee
Here we have used the Yang-Baxter equation (\ref{YangBaxter}).
\be
\cdots &=& \SSS_{1}(\theta_{1} \,; \xi_{-})
S_{1 2}(\theta_{1} + \theta_{2}) 
\tr_{0} \{ S_{0 1}(2\theta_{1}) ({\cal P}_{0 1} {\cal P}_{0 1})
S_{0 2}(\theta_{1} - \theta_{2}) {\cal P}_{0 1} 
\SSS_{0}(-\theta_{1} + i \pi\,; \xi_{+})^{t_{0}}\} \non \\
&=& \SSS_{1}(\theta_{1} \,; \xi_{-})
S_{1 2}(\theta_{1} + \theta_{2}) 
\tr_{0} \{ S_{0 1}(2\theta_{1}) {\cal P}_{0 1} 
\SSS_{0}(-\theta_{1} + i \pi\,; \xi_{+})^{t_{0}} \} 
S_{1 2}(\theta_{1} - \theta_{2}) \non \\
&=& \SSS_{1}(\theta_{1} \,; \xi_{-})
S_{1 2}(\theta_{1} + \theta_{2}) \SSS_{1}(\theta_{1} \,; \xi_{+})  
S_{1 2}(\theta_{1} - \theta_{2}) \,.
\ee
In passing to the last line, we have used the boundary cross-unitarity
relation (\ref{boundarycrossing}) with $\theta= {i\pi\over 2} -
\theta_{1}$, and the crossing relation $S_{01}(i\pi - \theta)^{t_{1}}
= S_{01}(\theta)$.  Comparing the last line to the expression
(\ref{openYangMatrix}) for the Yang matrix, we conclude that
\be
\tau(\theta_{1} | \theta_{1} \,, \theta_{2}) = Y_{(1)} \,.
\ee
For higher values of $N$, the proof is similar.

\section{Derivation of inversion identity}

In Section 4.1, we give the important inversion identity
(\ref{inversionidentity1}) - (\ref{functionf2}). Here we explain in 
more detail how we derived it. As already mentioned in text, the main 
idea is to formulate the fusion formula, following Ref. \cite{MN}, to 
which we shall refer as I. \footnote{In order to facilitate comparison 
with \cite{MN}, we use here similar notations.}

Although the ``dressed'' bulk $S$ matrix $S(\theta)$ (\ref{bulkS}) is
regular at $\theta=0$, the ``bare'' bulk $S$ matrix $R(\theta)$
(\ref{bulkRmatrix}) has a pole there.  In order to avoid complications
from this spurious pole, in this Appendix we rescale $R(\theta)$ by
the factor $\sinh \theta$; i.e., we take $R(\theta)$ to be given still
by (\ref{bulkRmatrix}), but now with matrix elements
\be
a_{\pm}(\theta) &=& \pm \sinh \theta - 2 i \sin B \pi \,, \qquad
b(\theta) = \sinh \theta \,, \non \\
c(\theta) &=& -2 i \sin B \pi\  \cosh {\theta\over 2} \,, \qquad
d(\theta) = -2 \sin B \pi\ \sinh {\theta\over 2} \,.
\ee
Keeping in mind the symmetries (\ref{Rsymmetries}) of the $R$ matrix, 
the unitarity relation (I 2.3) is
\be
R_{12}(\theta)\  R_{12}(-\theta) = \zeta (\theta) \id \,, \qquad 
\zeta (\theta) = - 4 \cosh^{2}{\theta\over 2}
(\sinh^{2}{\theta\over 2} + \sin^{2} B \pi) \,,
\ee 
and the crossing relation (I 2.4) is
\be
R_{12}(\theta) = V_1 \ R_{12}(-\theta - \rho)^{t_2}\  V_1 \,,
\ee 
with \footnote{Alternatively, choosing $\rho= -i\pi$, one has $V = \id$.}
\be
\rho = i \pi \,, \qquad V = \left( \begin{array}{rr}
  1  &0   \\ 
  0  &-1
\end{array} \right) \,.
\ee 
The matrix $R_{12}(\theta)$ at $\theta=-\rho$ is proportional to the 
one-dimensional projector $\tilde P_{12}^{-}$
\be
\tilde P_{12}^{-} = {1\over 2}  \left( \begin{array}{rrrr}
  1  &0  &0  &-1  \\ 
  0  &0  &0  &0  \\
  0  &0  &0  &0 \\
 -1  &0  &0  &1 
\end{array} \right) \,, \qquad 
(\tilde P_{12}^{-})^{2} = \tilde P_{12}^{-} \,.
\ee 
As explained in I, from the corresponding degeneration of the
(boundary) Yang-Baxter equation, one can derive identities which allow
one to prove that {\it fused} (boundary) $S$ matrices satisfy {\it
generalized} (boundary) Yang-Baxter equations.

The fused $R$ matrix is given by (I 2.13)
\be
R_{<12> 3}(\theta) = \tilde P_{12}^+\ R_{13}(\theta)\ 
R_{23}(\theta + \rho)\ \tilde P_{12}^+ \,,
\ee 
where $\tilde P_{12}^{+} = \id  - \tilde P_{12}^{-}$. An important 
observation (which one can verify by direct calculation) is that the 
fused $R$ matrix can be brought to upper triangular form by a 
similarity transformation \footnote{This observation is similar to, 
but not the same as, the one made by Felderhof \cite{Fe}. Indeed, in 
our language, he shows that $R_{13}(\theta) R_{23}(\theta + \rho)$ 
(i.e., the expression for the fused transfer matrix {\it without} the 
projectors $\tilde P_{12}^{+}$) can be brought to triangular form by 
a (somewhat more complicated) $\theta$-independent similarity 
transformation. Although for the case of periodic boundary conditions 
both approaches lead to the inversion identity, this appears to be no 
longer true for the case of boundaries.}
\be
X_{12}\ R_{<12> 3}(\theta)\ X_{12} = \mbox{ upper triangular }\,,
\ee
where the $4 \times 4$ matrix $X$ is independent of $\theta$, and is 
given by
\be
X = \left( \begin{array}{cccc}
  {1\over \sqrt{2}}  &0  &0  &{1\over \sqrt{2}} \\ 
  0  &-\sin {B \pi\over 2}  &\cos {B \pi\over 2}  &0  \\
  0  &\cos {B \pi\over 2}  &\sin {B \pi\over 2}  &0 \\
 {1\over \sqrt{2}}  &0  &0  &-{1\over \sqrt{2}} 
\end{array} \right) \,, \qquad 
X^{2} = \id \,.
\ee 
It follows that the fused monodromy matrices \footnote{For 
simplicity, we consider here the homogeneous case ($\theta_{i} =0 \,,
\quad i = 1 \,, \ldots \,, N$).} (I 4.7), (I 5.4), (I 5.5)
\be
T_{<12>}(\theta) &=& R_{<12> N}(\theta) \cdots R_{<12> 1}(\theta) \,, 
\non \\
\hat T_{<12>}(\theta + \rho) &=& R_{<12> 1}(\theta) \cdots 
R_{<12> N}(\theta) \,,
\ee
also become triangular by the same transformation.

Denoting (as in I) our ``bare'' boundary $S$ matrices 
$\R(\theta \,; \varphi_{-} )$,  
$\R(-\theta + i \pi \,; \varphi_{+} )$ by 
$K^{-}(\theta)$, $K^{+}(\theta)$, respectively, the corresponding 
fused matrices are given by (I 3.5), (I 3.9)
\be
K^-_{<12>}(\theta) &=& \tilde P_{12}^+\ K^-_1(\theta)\ 
R_{12} (2\theta + \rho)\ K^-_2(\theta + \rho)\ \tilde P_{12}^+ 
\,, \non \\
K^+_{<12>}(\theta) &=&
\{ \tilde P_{12}^+\ K^+_1(\theta)^{t_1}\  R_{12} (-2\theta -3\rho)\ 
K^+_2(\theta + \rho)^{t_2} \tilde P_{12}^+ \}^{t_{12}} \,,
\ee 
since $M = V^{t} V = \id$.

Remarkably, the fused $K$ matrices are also brought to upper 
triangular form by the {\it same} similarity transformation
\be
X_{12}\ K_{<12>}^{\mp}(\theta)\ X_{12} = \mbox{ upper triangular }\,.
\ee
It follows that the fused transfer matrix $\tilde \ttt(\theta)$, which 
is given by (I 4.5), (I 4.6)
\be
\tilde \ttt(\theta) = \tr_{12}\ K^+_{<12>}(\theta)\ 
T_{<12>}(\theta)\ K^-_{<12>}(\theta)\ 
\hat T_{<12>}(\theta + \rho) \,,
\ee 
is proportional to the identity matrix,
\be
\tilde \ttt(\theta) \propto \id \,,
\ee
where the proportionality factor is determined from the diagonal 
elements of the various triangular matrices.

The fusion formula is given by (I 4.17), (I 5.1)
\be
\ttt(\theta)\ \ttt(\theta+\rho) = {1\over \zeta( 2\theta + 2\rho)} 
\left[ \tilde \ttt(\theta) + 
\Delta \left\{ K^+(\theta) \right\} \Delta \left\{ K^-(\theta) \right\}
\delta \left\{ T(\theta) \right\} \delta \left\{ \hat T(\theta) \right\} 
\right] \,, 
\label{fusion}
\ee 
where the transfer matrix $\ttt(\theta)$ is given 
by (\ref{baretransfer}) (see also (I 4.1), (I 4.2)), and the 
quantum determinants \cite{IK} are given by
(I 4.15), (I 5.3), (I 5.7)
\be
\delta \left\{ T(\theta) \right \} &=& \delta \left\{\hat T(\theta) \right\} 
= \zeta(\theta + \rho)^N \,, \non \\
\Delta \left\{ K^-(\theta) \right\} &=& 
\tr_{12} \left\{  \tilde P_{12}^- \ K^-_1(\theta)\ 
R_{12}(2\theta + \rho)\ K^-_2(\theta+\rho)\ V_1\ V_2\ \right\} \,, \non \\
\Delta \left\{ K^+(\theta) \right\} &=& 
\tr_{12} \left\{ \tilde P_{12}^- \ V_1\ V_2\ K^+_2 (\theta + \rho)\ 
R_{12}(-2\theta-3\rho)\  K^+_1(\theta) \right\} \,.
\ee
Reverting to the original normalization of the $R$ matrix by rescaling 
each of the transfer matrices $\ttt(\theta)$ in (\ref{fusion}) by $(\sinh 
\theta)^{- 2N}$, introducing the inhomogeneities $\theta_{i}$ in the 
obvious way, and factoring the result into a product of two factors, 
we arrive at the results 
(\ref{inversionidentity1}) - (\ref{functionf2}). \footnote{We have 
refrained from giving explicit results for the intermediate steps, 
which are rather unwieldy and not very illuminating. We have done 
these computations with the help of Mathematica.}


\begin{thebibliography}{99}
    
\bibitem{DF}
P. Di Vecchia and S. Ferrara, Nucl. Phys. {\it B130} (1977) 93.

\bibitem{Hr}
J. Hruby, Nucl. Phys. {\it B131} (1977) 275.

\bibitem{FGS}
S. Ferrara, L. Girardello and S. Sciuto, Phys. Lett. {\it B76} (1978) 
303.

\bibitem{GS}
L. Girardello and S. Sciuto, Phys. Lett. {\it B77} (1978) 267.

\bibitem{SW}
R. Shankar and E. Witten, Phys. Rev. {\it D17} (1978) 2134.

\bibitem{Ah1}
C. Ahn, Nucl. Phys. {\it B422} (1994) 449.

\bibitem{ZZ1}
A.B. Zamolodchikov and Al.B. Zamolodchikov, Ann. Phys. {\it 120} (1979) 253;
A.B. Zamolodchikov, Sov. Sci. Rev. {\it A2} (1980) 1.

\bibitem{GZ}
S. Ghoshal and A.B. Zamolodchikov, Int. J. Mod. Phys. {\it A9} (1994)
3841.

\bibitem{IOZ}
T. Inami, S. Odake and Y-Z Zhang, Phys. Lett. {\it B359} (1995) 118.

\bibitem{AK}
C. Ahn and W.M. Koo, J. Phys. {\it A29} (1996) 5845;
Nucl. Phys. {\it B482} (1996) 675.

\bibitem{MS1}
M. Moriconi and K. Schoutens, Nucl. Phys. {\it B487} (1997) 756.

\bibitem{Sa}
H. Saleur, 1998 Les Houches lectures, cond-mat/9812110.

\bibitem{YY} 
C.N. Yang and C.P. Yang, J. Math. Phys. {\it 10} (1969) 1115.

\bibitem{Za1}
Al.B. Zamolodchikov, Nucl. Phys. {\it B342} (1990) 695.

\bibitem{Za2}
Al.B. Zamolodchikov, Nucl. Phys. {\it B358} (1991) 497.

\bibitem{ZZ2}
A.B. Zamolodchikov and Al.B. Zamolodchikov, Nucl. Phys. {\it B379} 
(1992) 602.

\bibitem{LMSS}
A. LeClair, G. Mussardo, H. Saleur and S. Skorik, Nucl. Phys. {\it 
B453}(1995) 581.

\bibitem{BPZ}
A.A. Belavin, A.M. Polyakov and A.B. Zamolodchikov,
Nucl. Phys. {\it B241} (1984) 333.

\bibitem{Ca}
J. Cardy, Nucl. Phys. {\it B324} (1989) 581.

\bibitem{AL}
I. Affleck and A.W.W. Ludwig, Phys. Rev. Lett. {\it 67} (1991) 161.

\bibitem{BKT}
M. Bershadsky, V. Knizhnik and M. Teilman, Phys. Lett. {\it B 151} 
(1985) 31.

\bibitem{FQS}
D. Friedan, Z. Qiu and S.H. Shenker, Phys. Lett. {\it B 151} (1985) 37.

\bibitem{Za4}
Al.B. Zamolodchikov, ``Resonance factorized scattering and roaming 
trajectories,'' unpublished preprint (1991).

\bibitem{Ma}
M.J. Martins,  Phys. Lett. {\it B304} (1993) 111.

\bibitem{DR}
P. Dorey and F. Ravanini, Nucl. Phys. B406 (1993) 708.

\bibitem{LSS}
F. Lesage, H. Saleur and P. Simonetti, Phys. Lett. {\it B427} (1998) 85.

\bibitem{AR}
C. Ahn and C. Rim, J. Phys. {\it A32} (1999) 2509.

\bibitem{Mu}
G. Mussardo, Nucl. Phys. {\it B532} (1998) 529.

\bibitem{Ya}
C.N. Yang, Phys. Rev. Lett. {\it 19} (1967) 1312.

\bibitem{FS}
P. Fendley and H. Saleur, Nucl. Phys. {\it B428} (1994) 681.

\bibitem{GMN}
M.  Grisaru, L. Mezincescu and R.I. Nepomechie, J.  Phys.  {\it A28} 
(1995) 1027

\bibitem{MN}
L. Mezincescu and R.I. Nepomechie, J.  Phys.  {\it A25} 
(1992) 2533.

\bibitem{FW}
C. Fan and F.Y. Wu, Phys. Rev. {\it B2} (1970) 723.

\bibitem{Fe}
B.U. Felderhof, Physica {\it 65} (1973) 421; {\it 66} (1973) 279, 509.

\bibitem{ABL}
C. Ahn, D. Bernard and A. LeClair, Nucl. Phys. {\it B346} (1990) 409.

\bibitem{BL}
D. Bernard and A. LeClair, Commun. Math. Phys. {\it 142} (1991) 99.

\bibitem{Ah2}
C. Ahn, Nucl. Phys. {\it B354} (1991) 57.

\bibitem{VG}
S.N. Vergeles and V.M. Gryanik, Yad. Fiz. {\it 23} (1976) 1324.

\bibitem{STW}
B. Schroer, T.T. Truong and P.H. Weisz, Phys. Lett. {\it 63B} (1976) 
422.

\bibitem{ArKo}
I.Ya. Aref'eva and V.E. Korepin, JETP Lett. {\it 20} (1974) 312.

\bibitem{KS}
P.P. Kulish and E.K. Sklyanin, in {\it Lecture Notes in Physics}, v. 151,
(Springer, 1982) 61.

\bibitem{KBI}
V.E. Korepin, N.M. Bogoliubov, and A.G. Izergin, {\it Quantum Inverse 
Scattering Method, Correlation Functions and Algebraic Bethe Ansatz}
(Cambridge University Press, 1993).

\bibitem{Ne}
R.I. Nepomechie, Int. J. Mod. Phys. {\it B13} (1999) 2973.

\bibitem{Gh}
S. Ghoshal, Int. J. Mod. Phys. {\it A9} (1994) 4801.

\bibitem{Ch} 
I.V. Cherednik, Theor. Math. Phys. {\it 61} (1984) 977.

\bibitem{Sk}
E.K. Sklyanin, J. Phys. {\it A21} (1988) 2375.

\bibitem{Ba}
R.J. Baxter, {\it Exactly Solved Models in Statistical Mechanics}
(Academic Press, 1982).
    
\bibitem{Ka}
M. Karowski, Nucl. Phys. {\it B153} (1979) 244.

\bibitem{KRS}
P.P. Kulish, N.Yu. Reshetikhin and E.K. Sklyanin, 
Lett. Math. Phys. {\it 5} (1981) 393.

\bibitem{IK}
A.G. Izergin and V.E. Korepin, Sov. Phys. Doklady {\it 26} (1981) 653;
Nucl. Phys. {\it B205} (1982) 401.

\bibitem{DRTW}
P. Dorey, I. Runkel, R. Tateo and G. Watts, hep-th/9909216

\bibitem{Sc}
K. Schoutens, Nucl.Phys. {\it B344} (1990) 665.

\bibitem{Ah3}
C. Ahn, Prog. Theor. Phys. {\it 118} (1995) 165.

\bibitem{MS2}
M. Moriconi and K. Schoutens, Nucl. Phys. {\it B464} (1996) 472.

\bibitem{FI}
P. Fendley and K. Intriligator, Nucl. Phys. {\it B380} (1992) 265.

\bibitem{AN}
C. Ahn and R.I. Nepomechie, in preparation.

\end{thebibliography}
\end{document}